\documentclass[preprint,
superscriptaddress,
amsmath,
amssymb,
aps]{revtex4-1}

\usepackage{graphicx}
\graphicspath{{figures/}}
\usepackage{graphicx}
\graphicspath{{figures/}}
\usepackage[bookmarks=false]{hyperref} 
\usepackage{xcolor}
\hypersetup{
    colorlinks,
    linkcolor={red},
    citecolor={green},
    linkcolor={red}
}
\usepackage{lineno}
\usepackage[margin=1in]{geometry}
\usepackage[normalem]{ulem}

\newcommand{\mean}[1]{\left\langle #1 \right\rangle}
\newcommand{\epr}{\dot{S}}
\newcommand{\eprfield}{\dot{s}}
\newcommand{\epf}{\mathcal{E}}
\newcommand{\epffield}{\mathcal{E}}
\newcommand{\raw}[1]{\widetilde{#1}}
\newcommand{\smooth}[1]{\hat{#1}}
\newcommand{\eq}[1]{Eq.~\ref{#1}}
\newcommand{\fig}[1]{Fig.~\ref{#1}}
\newcommand{\sfig}[1]{Supplementary~Figure~\ref{#1}}
\newcommand{\nocontentsline}[3]{}
\newcommand{\tocless}[2]{\bgroup\let\addcontentsline=\nocontentsline#1{#2}\egroup}

\begin{document}

\title{Irreversibility in dynamical phases and transitions}

\author{Daniel S. Seara}
    \email{daniel.seara@yale.edu}
    \affiliation{Department of Physics, Yale University, New Haven, Connecticut 06511, USA}
    \affiliation{Systems Biology Institute, Yale University, West Haven, Connecticut 06516, USA}
\author{Benjamin B. Machta}
    \email{benjamin.machta@yale.edu}
    \affiliation{Department of Physics, Yale University, New Haven, Connecticut 06511, USA}
    \affiliation{Systems Biology Institute, Yale University, West Haven, Connecticut 06516, USA}
\author{Michael P. Murrell}
    \email{michael.murrell@yale.edu}
    \affiliation{Department of Physics, Yale University, New Haven, Connecticut 06511, USA}
    \affiliation{Systems Biology Institute, Yale University, West Haven, Connecticut 06516, USA}
    \affiliation{Department of Biomedical Engineering, Yale University, New Haven, Connecticut 06511, USA}

\begin{abstract}
Living and non-living active matter consumes energy at the microscopic scale to drive emergent, macroscopic behavior including traveling waves and coherent oscillations. Recent work has characterized non-equilibrium systems by their total energy dissipation, but little has been said about how dissipation manifests in distinct spatiotemporal patterns. We introduce a novel measure of irreversibility we term the \textit{entropy production factor} (EPF) to quantify how time reversal symmetry is broken in field theories across scales. We use the EPF to characterize a dynamical phase transition in simulations of the Brusselator, a prototypical biochemically motivated non-linear oscillator. The EPF quantifies the distribution of irreversibility across spatiotemporal frequencies as the Brusselator transitions from local to global coherent oscillations, bounding the energetic cost to establish spatially synchronized biochemical oscillations.
\end{abstract}
\maketitle


\tocless\section{\label{sec:intro} Introduction}
    In many-body systems, collective behavior that breaks time-reversal symmetry can emerge due to the consumption of energy by the individual constituents~\cite{Toner1998FlocksFlocking, Ramaswamy2010, Marchetti2013}. In biological, engineered, and other naturally out of equilibrium processes, entropy must be produced so as to bias the system in a ``forward" direction~\cite{Onsager1953, Parrondo2009EntropyTime, Seifert2012, Barato2015ThermodynamicProcesses, Machta2015, Bryant2019EnergySystems}. This microscopic breaking of time reversal symmetry can manifest at different length and time scales in different ways. For example, bulk order parameters in complex reactions can switch from exhibiting incoherent, disordered behavior to stable static patterns~\cite{Turing1990, Ouyang1991TransitionPatterns} or traveling waves of excitation~\cite{Prigogine1967IntroductionProcesses, Cross1993PatternEquilibrium} that break time reversal symmetry in both time and space simply by altering the strength of the microscopic driving force. Recent advances in stochastic thermodynamics have highlighted entropy production as a quantity to measure a system's distance from equilibrium~\cite{Roldan2010EstimatingTrajectories, Fodor2016HowMatter, Seara2018, Li2019QuantifyingCurrents, Frishman2018LearningTrajectories, Martinez2019InferringCurrents}. While much work has been done investigating the critical behavior of entropy production at continuous and discontinuous phase transitions~\cite{Gaspard2004FluctuationReactions, Tome2012EntropyStates, Noa2019EntropyTransitions, Zhang2016, Ge2009, Ge2011, Nguyen2018, Tociu2019}, dynamical phase transitions in spatially extended systems have only recently been investigated, and to date no non-analytic behavior in the entropy production has been observed~\cite{Falasco2018, Rana2020}.

    To address this, we introduce what we term the entropy production factor (EPF), a dimensionless function of frequency and wavevector that measures time reversal symmetry breaking in a system's spatial and temporal dynamics. The EPF is a strictly non-negative quantity that is identically zero at equilibrium, quantifying how far individual modes are from equilibrium. Integrating the EPF produces a lower bound on the entropy production rate (EPR) of a system. We illustrate how to calculate the EPF directly from data using the analytically tractable example of Gaussian fields obeying partly relaxational dynamics supplemented with out of equilibrium coupling~\cite{Hohenberg1977TheoryPhenomena}. We then turn to the Brusselator reaction-diffusion model for spatiotemporal biochemical oscillations to study the connections between pattern formation and irreversibility. As the Brusselator undergoes a Hopf bifurcation far from equilibrium, it's behavior transitions from incoherent and localized to coordinated and system-spanning oscillations in a discontinuous transition. The EPF quantifies the shift in irreversibility from high to low wave-number as this transition occurs, but the EPR is indistinguishable from that of the well-mixed Brusselator where synchronization cannot occur. Importantly, the EPF can be calculated in any number of spatial dimensions, making it broadly applicable to a wide variety of data types, from particle tracking to 3+1 dimensional microscopy time series.

\tocless\section{\label{sec:derivation} Entropy production factor derivation}
    Consider a system described by a set of $M$ real, random variables obeying some possibly unknown dynamics. A specific trajectory of the system over a total time $T$ is given by $\mathbf{X} = \lbrace X^i(t) | t \in [0, T] \rbrace$. Given an ensemble of trajectories, the average EPR, $\epr \equiv \mean{dS/dt}$, is bounded by~\cite{Kawai2007Dissipation:Perspective, Parrondo2009EntropyTime, Seifert2012}
    \begin{equation}
        \label{eq:eprDef}
        \epr \geq \lim_{T \to \infty} \dfrac{1}{T} D_{KL} \left(P[\mathbf{X}] \ \middle \| \ \widetilde{P}[\mathbf{X}] \right);\qquad
        D_{KL} \left(P[\mathbf{X}] \ \middle \| \ \widetilde{P}[\mathbf{X}] \right) = \left<\log\left(\frac{P[\mathbf{X}]}{\widetilde{P}[\mathbf{X}]}\right) \right>_{P[\mathbf{X}]}
    \end{equation}
    where we have set $k_B=1$ throughout and $D_{KL}$ denotes the Kullback-Leibler divergence which measures the distinguishability between two probability distributions. $P\left[ \mathbf{X} \right]$ and $\widetilde{P} [\mathbf{X}]$ are the steady state probability distribution functionals of observing the path $\mathbf{X}(t)$ of length $T$ and the probability of observing its reverse path, respectively. Therefore, the KL divergence in~\eq{eq:eprDef} measures the statistical irreversibility of a signal, and saturates the bound when $\mathbf{X}$ contains all relevant, non-equilibrium degrees of freedom.

    We further bound the irreversibility itself by assuming the paths obey a Gaussian distribution. Writing the Fourier transform of $X^i(t)$ as $x^i(\omega)$ and writing the column vector $\mathbf{x}(\omega) = \left( x^1(\omega), x^2(\omega), \ldots \right)^T$:
    \begin{equation}
        \label{eq:pathProbRV}
        P [\mathbf{x}(\omega)] = \dfrac{1}{Z} \displaystyle\prod_{\omega_n} \exp \left( -\dfrac{1}{2T} \mathbf{x}^\dagger C^{-1} \mathbf{x} \right),
    \end{equation}
    where $\mathbf{x}^\dagger$ denotes the conjugate transpose of the vector $\mathbf{x}$ evaluated at the discrete frequencies $\omega_n=2\pi n/T$. $C(\omega_n)$ is the covariance matrix in Fourier space with elements $C^{ij}(\omega_n) = \mean{x^i(\omega_n) x^j(-\omega_n)} / T$, and $Z$ is the partition function. The expression for $\widetilde{P}[\mathbf{x}]$ is identical but with $C^{-1}(\omega_n) \to C^{-1}(-\omega_n)$ (see Supplementary Material). Combining~\eq{eq:eprDef} with~\eq{eq:pathProbRV} and taking $T \to \infty$, we arrive at our main result:
    \begin{equation}
        \label{eq:spectralEPR_particles}
        \epr = \int \dfrac{d\omega}{2 \pi} \ \epf(\omega); \qquad \epf(\omega) =  \dfrac{1}{2} \left[ C^{-1} (-\omega)  - C^{-1} (\omega) \right]_{ij} C^{ji}(\omega).
    \end{equation}

    This defines the EPF, $\epf(\omega)$, which measures time reversal symmetry breaking interactions between $M \geq 2$ variables, while integrating $\epf$ gives $\epr$. $\epf(\omega)=D_{KL}(P[\mathbf{x}(\omega)] \ || \ P[\mathbf{x}(-\omega)])$ measures the Kullback-Leibler divergence between the joint distribution of $M$ modes at a single frequency $\omega$.  While this quantity does not scale with trajectory length, the density of modes near a particular frequency is related to the total trajectory time by $1/\Delta \omega =T/2\pi$. Since $\pm \omega$ modes must be complex conjugates of each other and an overall average phase is prohibited by time translation invariance, asymmetry between these distributions can only be captured by relative phase relationships, quantified by their correlation functions. $\epf$ is large when one variable tends to lead another in phase, implying a directed rotation between these variables in the time domain.

    As mentioned above, $P[\mathbf{x}(\omega)]$ describes the dynamics of a non-equilibrium steady state, and no reversal of external protocol is assumed. Further, in writing an expression for $\widetilde{P}[\mathbf{x}(\omega)]$, we assume that the observables are scalar, time-reversal symmetric quantities, such as the chemical concentrations we analyze below.

    The Gaussian assumption we make here makes~\eq{eq:spectralEPR_particles} exact only for systems obeying linear dynamics. Nevertheless, $\epf$ is still defined for non-linear systems, where the integrated $\epf$ lower bounds the true $\epr$.  To see this, consider projecting complex dynamics onto Gaussian dynamics by choosing a data processing procedure which preserves two point correlations but which removes higher ones.  This can be accomplished by multiplying every frequency by an independent random phase --- a post processing procedure which can be applied to individual trajectories.  Post-convolution, the integrated EPF is equal to the KL divergence rate between forward and backwards rates. From the data processing inequality, the KL divergence rate of the true fields must be higher, so that the integrated EPF lower bounds the true entropy production rate. [See Supplementary Material for a more detailed derivation.]  In addition to bounding the true $\epr$, we expect the integral of $\epf$ to be a good approximation for the wide class of systems where linearization is reasonable. Such Gaussian approximations are starting points in many field theories, with higher order interactions accounted for by adding anharmonic terms in the action of~\eq{eq:pathProbRV}. While this is not our focus here, we expect these additional terms to systematically capture corrections to $\epr$ that do not appear in~\eq{eq:spectralEPR_particles}. As $C^{ii}(\omega) = C^{ii}(-\omega)$, the only contributions to $\epf$ come from the cross-covariances between the random variables of interest. As such, this bound yields exactly 0 for a single variable even though higher order terms may contribute to $\epr$.

    This formulation extends naturally to random fields.  For $M$ random fields in $d$ spatial dimensions, $\boldsymbol{\phi} = \lbrace \phi^i(\mathbf{x}, t) | t \in [0,T], \mathbf{x} \in \mathbb{R}^d \rbrace$, the EPR density, $\eprfield \equiv \epr / V$ where $V$ is the system volume, is [See Supplementary Material]:
    \begin{equation}
        \label{eq:spectralEPR_fields}
        \eprfield = \int \dfrac{d \omega}{2 \pi} \dfrac{d^d \mathbf{q}}{(2 \pi)^d} \ \epffield(\mathbf{q}, \omega); \qquad \epffield(\mathbf{q}, \omega) =  \dfrac{1}{2} \left[ C^{-1} (\mathbf{q}, -\omega)  - C^{-1} (\mathbf{q}, \omega) \right]_{ij} C^{ji}(\mathbf{q}, \omega).
    \end{equation}
    where $C^{ij}(\mathbf{q},\omega)$ is the dynamic structure factor and $\epffield(\mathbf{q},\omega)$ is now a function of wavevector and frequency [see Supplementary Material]~\footnote{Here, we have neglected a term in $D_{KL}$ that depends on the partition function, which can only be non-zero in systems that lack both parity and rotational symmetry}.

    Even without an explicit, analytic expression for the structure factor, $C$, we can estimate $\epf$ from data. To use~\eq{eq:spectralEPR_fields}, we consider data of $N$ finite length trajectories of $M$ variables over a time $T$ in $d$ spatial dimensions. Each dimension has a length $L_i$. We create an estimate of the covariance matrix, $\raw{C}(\mathbf{q}, \omega)$, from time-series using standard methods [see Methods]. These measurements will inevitably contain noise that is not necessarily time-reversal symmetric, even for an equilibrium system. Noise due to thermal fluctuations and finite trajectory lengths in the estimate of $\raw{C}$ from a single experiment ($N=1$) will systematically bias our estimated $\epf$ by $\Delta \epf = M(M-1)/2$ at each frequency and will thereby introduce bias and variance in our measurement of $\eprfield$. We can simply remove the bias from our measured $\epffield$, but to reduce the variance, we smooth $\raw{C}$ by component-wise convolution with a multivariate Gaussian of width $\boldsymbol{\sigma} = (\sigma_{q_1}, \ldots, \sigma_{q_d}, \sigma_\omega)$ in frequency space, giving $\smooth{C}$. This is equivalent to multiplying each component of the time domain $\raw{C}(\mathbf{r}, t)$ by a Gaussian, cutting off the noisy tails in the real space covariance functions at large lag times. We then use $\smooth{C}$ in~\eq{eq:spectralEPR_fields} to create our final estimator for the EPF, $\smooth{\epffield}$, and thereby the EPR, $\smooth{\eprfield}$. We calculate and remove the bias in $\smooth{\epffield}$ and $\smooth{\eprfield}$ in all results below [see Methods]. Smoothing $\raw{C}$ with increasingly wide Gaussians in $\omega$ and $\mathbf{q}$ leads to a systematic decrease in $\smooth{\eprfield}$ due to reduced amplitudes in $\raw{C}$ (\sfig{sfig:eprPlot_bruss_varySigma}).

\tocless\section{\label{sec:results}Results}
    To illustrate the information contained in $\epffield$, its numerical estimation, and the accuracy of $\smooth{\eprfield}$, we analyze simulations of coupled, 1 dimensional Gaussian stochastic fields for which $\epffield$ and $\eprfield$ can be calculated analytically. We then study simulations of the reaction-diffusion Brusselator, a prototypical model for non-linear biochemical oscillators, and use $\epffield$ to study how irreversibility manifests at different time and length scales as the system undergoes a Hopf bifurcation~\cite{Strogatz2015NonlinearEngineering}.

    \tocless\subsection{\label{sec:gaussfield} Driven Gaussian fields}
        Consider two fields obeying Model A dynamics~\cite{Hohenberg1977TheoryPhenomena} with non-equilibrium driving parametrized by $\alpha$:
        \begin{align}
            \label{eq:gaussfield}
            \begin{split}
                \partial_t \phi(x, t) &= -D \dfrac{\delta \mathcal{F}}{\delta \phi} - \alpha \psi + \sqrt{2D} \xi_\psi \\
                \partial_t \psi(x, t) &= -D \dfrac{\delta \mathcal{F}}{\delta \psi} + \alpha \phi + \sqrt{2D} \xi_\phi,
            \end{split}
        \end{align}
        where $\xi(\mathbf{x}, t)$ is Gaussian white noise with variance $\mean{\xi^i(\mathbf{x}, t) \xi^j(\mathbf{x}', t')} = \delta^{ij} \delta(\mathbf{x} - \mathbf{x}') \delta(t-t')$, $D$ is a relaxation constant, and $\delta \mathcal{F}/ \delta \phi$ is the functional derivative with respect to $\phi$ of the free energy $\mathcal{F}$ given by:
        \begin{equation}
            \label{eq:modelAFreeEnergy}
            \mathcal{F} = \int dx \left[ \dfrac{r}{2} (\phi^2 + \psi^2) + \dfrac{1}{2} \left( |\partial_x \phi|^2 + |\partial_x \psi|^2 \right) \right],
        \end{equation}
        so that the fields have units of $\ell^{1/2}$ and $r$ penalizes large amplitudes.

        The EPR density, $\eprfield$, is calculated analytically in two ways. First, we solve~\eq{eq:eprDef} directly using the Onsager-Machlup functional for the path probability functional of $\boldsymbol{\eta}(x, t) = \left( \phi(x,t), \psi(x, t) \right)^T$~\cite{Onsager1953, Nardini2017}. Second, the covariance matrices are calculated analytically, used to find $\epffield$ through~\eq{eq:spectralEPR_fields}, and integrated to find $\eprfield$. Both cases give the same result for $\eprfield$. The result for both $\epffield$ and $\eprfield$ are [Supplementary Material]:
        \begin{equation}
        \label{eq:gaussfieldEPR}
            \epffield^\mathrm{DGF} = \dfrac{8 \alpha^2 \omega^2}{(\omega^2 - \omega_0^2(q))^2 + (2D(r + q^2)\omega)^2}, \quad
            \eprfield^\mathrm{DGF} = \dfrac{\alpha^2}{D\sqrt{r}}.
        \end{equation}

        We see that $\epf^\mathrm{DGF} \geq 0$ and exhibits a peak at $(q, \omega) = (0,\ \omega_0(0))$, where $\omega_0(q) = \sqrt{(D(r + q^2))^2 + \alpha^2}$, indicating that the system is driven at all length scales with a driving frequency of $\alpha$, dampened by an effective spring constant $Dr$. In addition, it is clear that multiple combinations of $\alpha$, $r$, and $D$ can give the same value for $\eprfield$ while $\epffield$ distinguishes between equally dissipative trajectories in the shape and location of its peaks. In this way, $\epffield$ gives information about the form of the underlying dynamics not present in the total EPR. We note that $\epf^\mathrm{DGF}$ is also recovered using an appropriately modified version of the generalized Harada-Sasa Relation introduced in~\cite{Nardini2017} [see Supplementary Material].

        \begin{figure*}
            \centering
            \includegraphics[width=\textwidth]{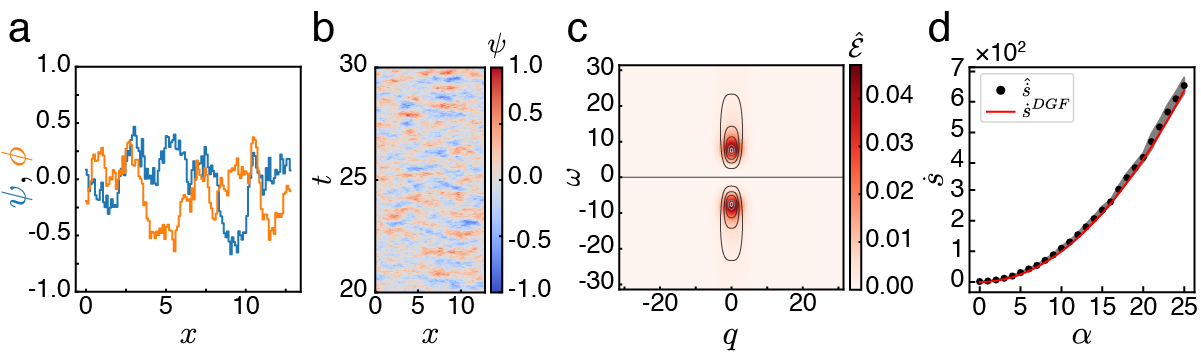}
            \caption{\textbf{Entropy production rate and entropy production factor are well estimated for driven Gaussian fields.} (a) Snapshot of typical configurations of both fields, $(\psi, \phi)$ obeying~\eq{eq:gaussfield} for $\alpha=7.5$. (b) Subsection of a typical trajectory for one field for $\alpha=7.5$ in dimensionless units. Colors indicate the value of the field at each point in spacetime. (c) $\smooth{\epffield}$ for $\alpha = 7.5$ averaged over $N= 10$ simulations. Contours show level sets of $\epffield^{DGF}$. (d) Measured $\eprfield$ vs. $\alpha$ for simulations of total time $T = 50$ and length $L=12.8$. Red line shows the theoretical value, $\eprfield^{DGF}$. Mean $\pm$ standard deviation of $\smooth{\eprfield}$ given by black dots and shaded area. See Supplementary Material for all simulation parameters.}
            \label{fig:epr_gaussfield}
        \end{figure*}

        We perform simulations to assess how well $\epffield$ can be extracted from time series data of fields [See methods for details]. The estimated $\smooth{\epffield}$ shows excellent agreement with~\eq{eq:gaussfieldEPR} (\fig{fig:epr_gaussfield}). Integrating $\smooth{\epffield}$ gives $\smooth{\eprfield}$, which also shows good agreement with $\eprfield^{DGF}$.

        Our estimator gives exact results for the driven Gaussian fields because the true path probability functional for these fields is Gaussian. In contrast, the complex patterns seen in nature arise from systems obeying highly non-linear dynamics. For such dynamics, our Gaussian approximation is no longer exact but provides a lower bound on the total irreversibility. To investigate how irreversibility correlates with pattern formation, we study simulations of the Brusselator model for biochemical oscillations~\cite{Prigogine1968SymmetryII}. We begin by describing the various dynamical phases of the equations of motion. Next, we calculate $\epffield$ and $\epr$ for only the reactions before adding diffusion to study the synchronized oscillations that arise in the 1 dimensional reaction-diffusion system.

    \tocless\subsection{\label{nonlinearSystems}Reaction-diffusion Brusselator}
\begin{figure*}
            \centering
            \includegraphics[width=\textwidth]{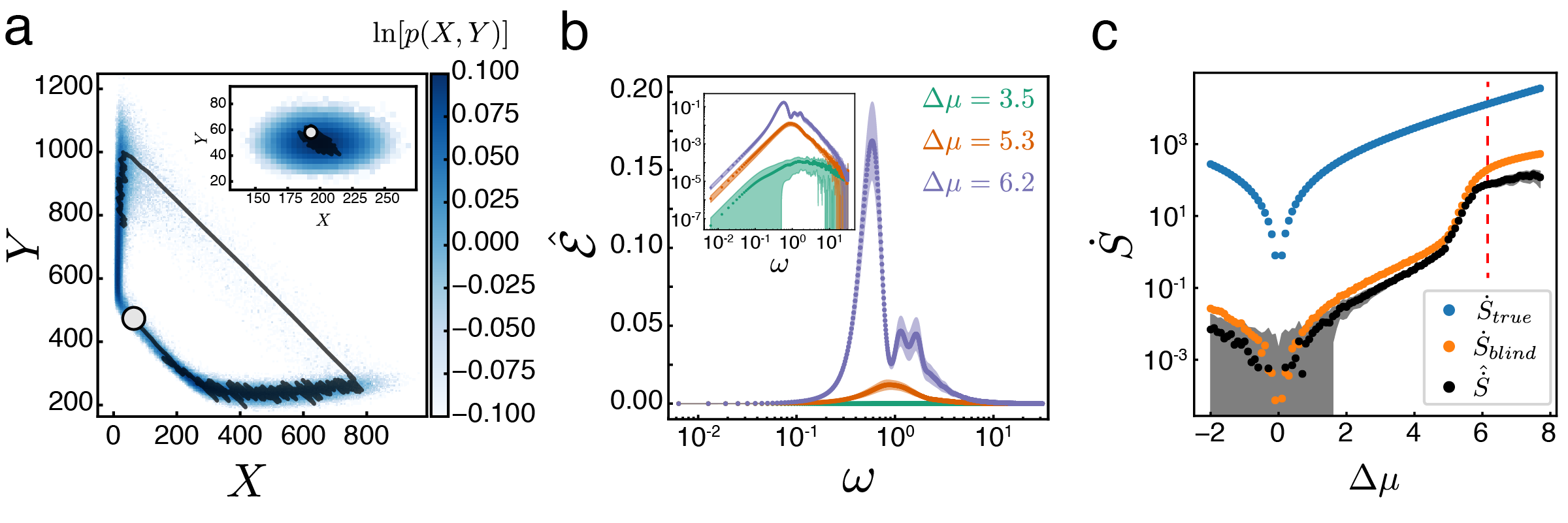}
            \caption{\textbf{$\epr$ and $\epf$ for well-mixed Brusselator.} (a) Typical trajectory in $(X,Y)$ space for $\Delta \mu=6.2$. The occupation probability distribution is shown in blue, with a subsection of a typical trajectory shown in black. The end of the trajectory is marked by the white circle. Inset shows the same information for the system at equilibrium, where $\Delta \mu = 0$, with the same colorbar as the main figure. (b) $\smooth{\epf}$ for $\Delta \mu = [3.5, 5.3, 6.2]$ shown in green, orange, and purple, respectively. Shaded area shows mean $\pm$ std of $\smooth{\epf}$ for $N=50$ simulations. $\smooth{\epf}$ is symmetric in $\omega$, so only the positive axis is shown. Inset shows the same curves on a log-log scale. (c) $\epr$ as a function of $\Delta \mu$. Blue, orange, and black points show results for $\epr_{\mathrm{true}}$, $\epr_{\mathrm{blind}}$, and $\smooth{\epr}$, respectively. Shaded area shows mean $\pm$ std of $\smooth{\epr}$ for $N=50$ simulations. Red dashed line indicates $\Delta \mu_{\mathrm{HB}}$. See Supplementary Material for all simulation parameters.}
            \label{fig:epr_brusselator}
        \end{figure*}

        We use a reversible Brusselator model~\cite{Prigogine1968SymmetryII, Qian2002ConcentrationSystem, Fei2018, Rana2020} with dynamics governed by the reaction equations:
        \begin{equation}
            \label{eq:brusselator}
            A \underset{k^-_1}{\overset{k^+_1}{\rightleftharpoons}} X; \quad
            B + X \underset{k^-_2}{\overset{k^+_2}{\rightleftharpoons}} Y + C; \quad
            2X + Y \underset{k^-_3}{\overset{k^+_3}{\rightleftharpoons}} 3X;
        \end{equation}
        where $\lbrace A,B,C \rbrace$ are external chemical baths with fixed concentrations $\lbrace a, b, c \rbrace$, and all the reactions occur in a volume $V$ (\fig{fig:epr_brusselator}a). The system is in equilibrium when the external chemical baths and reaction rates obey $B k_2^+ k_3^+ = C k_2^- k_3^-$. When this equality is violated, the system is driven away from equilibrium and exhibits cycles in the $(X, Y)$ plane. Defining
        \begin{equation}
            \label{eq:brussAlpha}
            \Delta \mu = \ln \left( \dfrac{B k_2^+ k_3^+}{C k_2^- k_3^-} \right),
        \end{equation}
        the Brusselator is at equilibrium when $\Delta \mu = 0$ and is driven into a non-equilibrium steady state when $\Delta \mu \neq 0$. We vary $B$ and $C$ to change $\Delta \mu$ while keeping the product $(b k_2^+ k_3^+)(c k_2^- k_3^-) = 1$, keeping the rate at which reactions occur constant for all $\Delta \mu$~\cite{Maes2006Time-SymmetricSystems}.

        As $\Delta \mu$ increases, the macroscopic version of~\eq{eq:brusselator} undergoes dynamical phase transitions. For all $\Delta \mu$, there exists a steady state $(X_{ss}, Y_{ss})$, the stability of which is determined by the relaxation matrix, $R$ [Supplementary Material]. The two eigenvalues of $R$ , $\lambda_\pm$, divide the steady state into four classes~\cite{Strogatz2015NonlinearEngineering}:
        \begin{enumerate}
            \item $\lambda_\pm \in \mathbb{R}_{<0} \to$ Stable attractor, no oscillations
            \item $\lambda_\pm \in \mathbb{C},\ \mathrm{Re}[\lambda_\pm] < 0 \to$ Stable focus
            \item $\lambda_\pm \in \mathbb{C},\ \mathrm{Re}[\lambda_\pm]
            > 0 \to$ Hopf Bifurcation, limit cycle
            \item $\lambda_\pm \in \mathbb{R}_{>0} \to$ Unstable repeller
        \end{enumerate}
        The eigenvalues undergo these changes as $\Delta \mu$ changes, allowing us to consider $\Delta \mu$ as a bifurcation parameter. We define $\Delta \mu_{\text{HB}}$ as the value of $\Delta \mu$ where the macroscopic system undergoes the Hopf bifurcation.

        Non-equilibrium steady states are traditionally characterized by their circulation in a phase space~\cite{Schnakenberg1976NetworkSystems, Zia2007, Battle2016, Gingrich2017, Weiss2019NonequilibriumSystem}. One may then question how it is possible to detect non-equilibrium effects in the Brusselator when the system's steady state is a stable attractor with no oscillatory component. While this is true for the macroscopic dynamics used to derive $\lambda_{\pm}$, we simulate a system with finite numbers of molecules subject to fluctuations. These stochastic fluctuations give rise to circulating dynamics, even when the deterministic dynamics do not~\cite{Qian2002ConcentrationSystem}. We see persistent circulation in the $(X, Y)$ plane when $\lambda_\pm \in \mathbb{R}_{<0}$, with the vorticity changing sign around $\Delta \mu = 0$ (\sfig{sfig:brusselator_evalTraj}).

        In order to assess the accuracy of our estimated EPR, $\smooth{\epr}$, we calculate an estimate of the true EPR, $\epr_{\text{true}}$, for a simulation of~\eq{eq:brusselator} by calculating the exact entropy produced by each reaction that occurs in the trajectory~\cite{Esposito2010ThreeTheorems}, and then fitting a line to the cumulative sum (\sfig{sfig:brussFit}, Methods). We find that $\smooth{\epr}$ significantly underestimates $\dot{S}_{\text{true}}$ (note the logged axes in \fig{fig:epr_brusselator}c) due to the Brusselator's hidden dynamics. In the Brusselator, information is lost because the observed trajectories are coarse-grained --- they do not distinguish between reactions that take place forward through the second reaction or backwards through the third reaction in~\eq{eq:brusselator}. These pathways would be distinguishable if trajectory of $B$ and $C$ were also observable. Our method relies purely on system dynamics to give $\smooth{\epr}$.~\eq{eq:eprDef} is true only if all microscopic details are captured by trajectories $\mathbf{X}$. If $\mathbf{X}$ is already coarse-grained, multiple microscopic trajectories will be indistinguishable and~\eq{eq:eprDef} will underestimate the true entropy production rate due to the data processing inequality ~\cite{Shiraishi2015FluctuationDynamics, Polettini2017EffectiveObserver, Martinez2019InferringCurrents}.

        In order to account for this, we recalculate $\epr$ by considering the rate at which a given transition can occur as the sum over all chemical reactions that give the same dynamics [see Methods]. For example, a transition from $(X, Y) \to (X-1, Y+1)$ can occur via reaction $k_2^+$ or $k_3^-$ in the Brusselator, each of which produces a different amount of entropy in general. Looking only in the $(X, Y)$ plane, it is impossible to tell which reaction took place. When calculating the entropy produced by only the observable dynamics, the rate of making the transition $(X, Y) \to (X-1, Y+1)$ is $k_f= k_2^++k_3^-$, while the rate of making the reverse transition is $k_r= k_2^-+k_3^+$, and the entropy produced is $\ln(k_f/k_r)$. This ``blinded" estimate of the EPR, $\epr_{\text{blind}}$, shows excellent agreement with $\smooth{\epr}$, indicating that the Gaussian approximation provides a good estimate for the observable dynamics even when the system is highly nonlinear.

        To further benchmark our estimator, we calculate $\epr$ using two alternative methods, one based on the thermodynamic uncertainty relation~\cite{Barato2015ThermodynamicProcesses, Horowitz2017, Li2019QuantifyingCurrents} and one based on measuring first passage times~\cite{Roldan2015}. These two other methods also approximate $\epr_\mathrm{blind}$ and provide a looser bound than $\smooth{\epr}$ (\sfig{sfig:eprPlot_bruss_altMethods}).

\begin{figure}
            \centering
            \includegraphics[width=\textwidth]{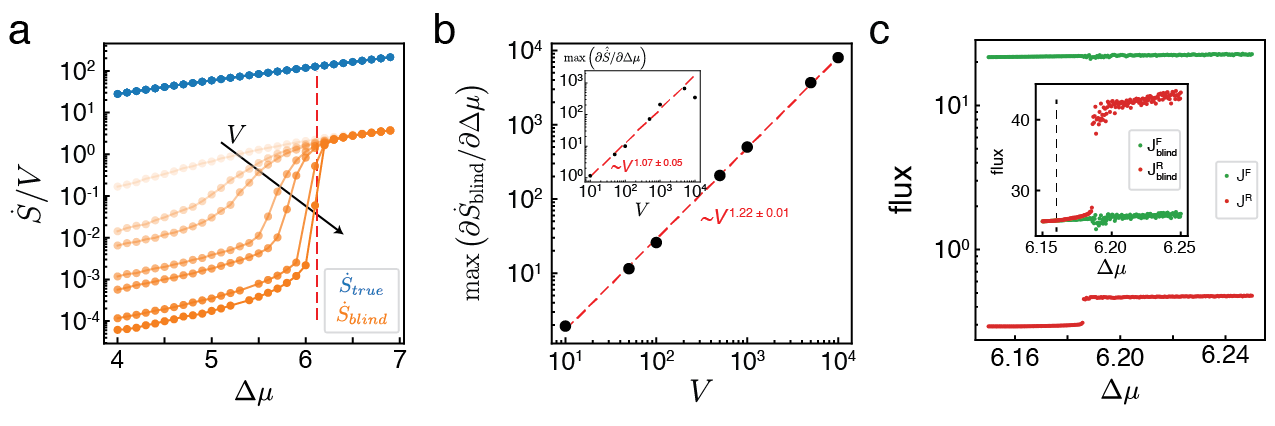}
            \caption{\textbf{Finite size scaling of $\epr$} (a) $\epr_\mathrm{true}/V$ (blue) and $\epr_\mathrm{blind}/V$ (orange) for system volumes $V=[10, 50, 100, 500, 1000, 5000, 10000]$, showing an increasingly sharp transition in $\epr_\mathrm{blind}$, but not in $\epr_\mathrm{true}$. $\epr_\mathrm{blind}$ shows no volume dependence below the transition, and is linear dependent on $V$ above it. (b) Maximum value of $\partial \epr_\mathrm{blind} / \partial \Delta \mu$ shows a power-law dependence with volume. Inset shows the same measurement for $\partial \smooth{\epr}/\partial \Delta \mu$. (c) Forward and reverse fluxes, $ J^\mathrm{F}$ (green) and $J^\mathrm{R}$ (red), obtained from numerical integration of deterministic equations of motion for the Brusselator. Inset shows $J^\mathrm{F}_\mathrm{blind}$ (green) and $J^\mathrm{R}_\mathrm{blind}$ (red)}
            \label{fig:bruss_eprScaling}
        \end{figure}

        Prior to $\Delta \mu_{\mathrm{HB}}$, both $\smooth{\epr}$ and $\epr_{\mathrm{blind}}$ show a shift in their trends, but $\epr_{\mathrm{true}}$ does not. The smooth transition is due to the finite system size we employ, and gets sharper as a power law as the system gets larger (\fig{fig:bruss_eprScaling}a). The power law exponent measured from $\smooth{\epr}$ is nearly linear, consistent with the Gaussian assumption. The exponent differs from that of $\epr_\mathrm{blind}$ because our Gaussian assumption breaks down at the high values of $\Delta \mu$ where the maximum slope occurs (\fig{fig:bruss_eprScaling}b).

        The Hopf bifurcation for the Brusselator is supercritical~\cite{Zhang2016}, meaning the limit cycle grows continuously from the fixed point when $\Delta \mu - \Delta \mu_\mathrm{HB} \ll 1$. Further from the critical point, the trajectory makes a discontinuous transition. At our resolution in $\Delta \mu$, this discontinuous transition is what underlies the shift in $\epr_\mathrm{blind}$ of the Brusselator. This same transition is present in $\epr_\mathrm{true}$, but is difficult to detect numerically for reasons we explain here. In the deterministic limit, $\epr_\mathrm{true} = \Delta \mu \left(J^\mathrm{F} - J^\mathrm{R}\right)$, where $J^\mathrm{F} = b \langle x \rangle k_2^+$ and $J^\mathrm{R} = c \langle y \rangle k_2^-$ are the forward and reverse fluxes for transforming a $B$ molecule into a $C$ molecule. $\langle x \rangle$ is a constant, but by numerically integrating the deterministic version for \eq{eq:brusselator}, we observe a discontinuity in $\langle y \rangle$ above the Hopf bifurcation. However, $J^\mathrm{F} \gg J^\mathrm{R}$, obscuring the discontinuity   in $\epr_\mathrm{true}$ (\fig{fig:bruss_eprScaling}c). Upon coarse-graining, we have $\epr_\mathrm{blind} = \Delta \mu \left( J^\mathrm{R}_\mathrm{blind} - J^\mathrm{F}_\mathrm{blind} \right)$, with $J^\mathrm{F}_\mathrm{blind} = b \langle x \rangle k_2^+ + \langle x \rangle^3 k_3^-$ and $J^\mathrm{R} = c \langle y \rangle k_2^- + \langle x \rangle^2 \langle y \rangle k_3^+$. These two terms are equal to each other for $\Delta \mu < \Delta \mu_\mathrm{HB}$ and diverge continuously when $\Delta \mu \gtrapprox \Delta \mu_\mathrm{HB}$, followed by the relatively large discontinuity in $J^\mathrm{R}_\mathrm{blind}$ (\fig{fig:bruss_eprScaling}c, inset).

        One gains further insight into the dynamics through the transition by studying $\smooth{\epf}$ (\fig{fig:epr_brusselator}b). For $\Delta \mu < \Delta \mu_{\text{HB}}$, $\smooth{\epf}$ exhibits a single peak that increases in amplitude while decreasing in frequency as $\Delta \mu$ increases. Above $\Delta \mu_{\text{HB}}$, the peak frequency makes a discontinuous jump, the magnitude of the peak grows rapidly, and additional peaks at integer multiples of the peak frequency appear due to the non-linear shape of the limit cycle attractor. These harmonics are expected for dynamics on a non-circular path. For $\Delta \mu < \Delta \mu_\mathrm{HB}$, the magnitude of the peak is independent of system volume, while it gains a linear volume dependence in the limit cycle. The width of the peak is also maximized near the transition, reflecting a superposition of frequencies present in the trajectories (\sfig{sfig:bruss_epfscaling}).

        To investigate how dynamical phase transitions manifest in the irreversibility of spatially extended systems, we simulate a reaction-diffusion Brusselator on a 1 dimensional periodic lattice with $L$ compartments, each with volume $V$, spaced a distance $h$ apart. The full set of reactions are now
        \begin{align}
            \label{eq:brussfield}
            \begin{split}
                A_i \underset{k^-_1}{\overset{k^+_1}{\rightleftharpoons}} X_i; \quad
                B_i + X_i \underset{k^-_2}{\overset{k^+_2}{\rightleftharpoons}} Y_i + C_i; \quad
                2X_i + Y_i \underset{k^-_3}{\overset{k^+_3}{\rightleftharpoons}} 3X_i; \\
                X_i \underset{d_X}{\overset{d_X}{\leftrightharpoons}} X_{i+1}; \qquad
                Y_i \underset{d_Y}{\overset{d_Y}{\leftrightharpoons}} Y_{i+1}; \qquad i \in [1, L]
            \end{split}
        \end{align}
        where $d_j = D_j / h^2$, and $D_j$ is the diffusion constant of chemical species $j = \lbrace X, Y \rbrace$. Qualitatively different dynamics occur based on the ratio $D_X/D_Y$. $D_X/D_Y \ll 1$ yields static Turing patterns \cite{Turing1990, Falasco2018}. We focus on the $D_X/D_Y \gg 1$ regime which exhibits dynamic, excitable waves. All values of $\{ a_i, b_i, c_i \}$ are kept constant in each compartment.

        \begin{figure*}
            \centering
            \includegraphics[width=\textwidth]{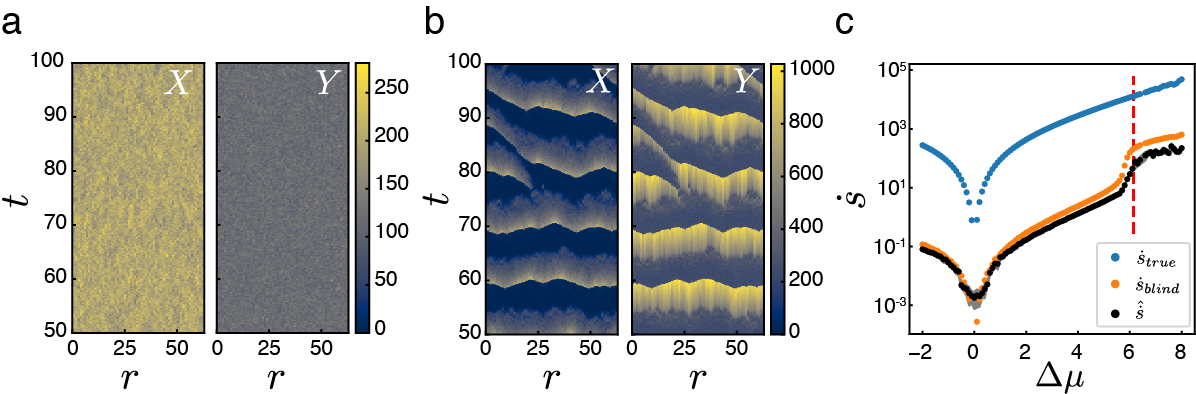}
            \caption{\textbf{1 dimensional reaction-diffusion Brusselator network shows emergent collective behavior above Hopf Bifurcation} (a) Subsection of a typical trajectory for $X(r, t)$ and $Y(r,t)$ for (a) $\Delta \mu = 3.5$, below the Hopf Bifurcation and (b) $\Delta \mu = 6.2$, above it. Color indicates the local number of the chemical species. (c) $\eprfield$ as a function of $\Delta \mu$. Blue, orange, and black points show results for $\epr_{\mathrm{true}}$, $\epr_{\mathrm{blind}}$, and $\smooth{\epr}$, respectively. Shaded area shows mean $\pm$ std of $N=10$ simulations.  Red dashed line indicates $\Delta \mu_{\mathrm{HB}}$. See Supplementary Material for all simulation parameters.}
            \label{fig:epr_brussfield}
        \end{figure*}

        In the steady state, the reaction-diffusion Brusselator has the same dynamics as the well mixed Brusselator, and so it is not surprising that it's EPR curve as a function of $\Delta \mu$ is similar (\sfig{sfig:bruss_vs_brussfield_EPR}). However, unlike the well-mixed system, the Hopf bifurcation signals the onset of qualitatively distinct dynamics in the reaction-diffusion system. Prior to the Hopf bifurcation, there are no coherent, spatial patterns in the system's dynamics (\fig{fig:epr_brussfield}a). Above the Hopf bifurcation, system-spanning waves begin to emerge that synchronize the oscillations across the system (\fig{fig:epr_brussfield}b).

\begin{figure*}
            \centering
            \includegraphics[width=\textwidth]{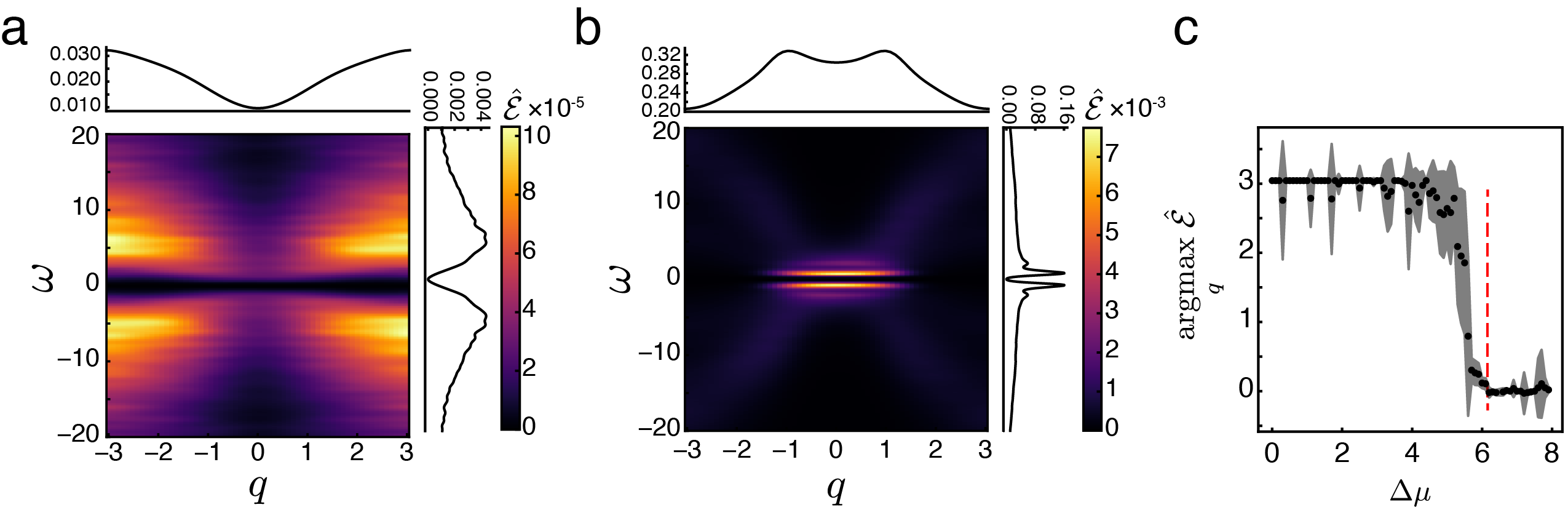}
            \caption{\textbf{Entropy production factor highlights macroscopic behavior after undergoing a Hopf bifurcation in the reaction-diffusion Brusselator model.} (a) $\mean{\epffield }$ over $N=10$ simulations for $\Delta \mu = 4.0$, i.e. $\Delta \mu < \Delta \mu_{\text{HB}}$. Line plots on top and left of figure show marginals over $\omega$ and $q$, respectively. (b) Similar to (a), but for  $\Delta \mu = 6.2$, i.e. $\Delta \mu > \Delta \mu_{\text{HB}}$. (c) Wavenumber, $q$, that maximizes $\smooth{\epffield}$ as a function of $\Delta \mu$. Red line shows $\Delta \mu_{\text{HB}}$.}
            \label{fig:epfPlot_brussfield}
        \end{figure*}

        Throughout these changes, the system is driven further from equilibrium, as reflected in the increasing $\smooth{\eprfield}$ (\fig{fig:epr_brussfield}c). The shift to collective behavior is not reflected in $\eprfield$ as it is almost identical to $\epr$ found for the well-mixed Brusselator (\sfig{sfig:bruss_vs_brussfield_EPR}). Instead, $\epffield$ carries the signature of the dynamical phase transition. For $\Delta \mu < \Delta \mu_{\text{HB}}$, $\smooth{\epffield}$ shows peaks at high wavenumbers, reflecting that irreversibility is occurring incoherently over short length scales. Above $\Delta \mu_{\text{HB}}$, as the system shows synchronized oscillations, there is an abrupt shift in the peaks of $\smooth{\epf}$ to low $q$, indicating that this collective behavior carries the majority of the irreversibility (\fig{fig:epfPlot_brussfield}b,c). We also infer that the collective behavior is partially composed of traveling waves due to the streaks in $\smooth{\epffield}$ (\fig{fig:epfPlot_brussfield}b). The slight offset in the transition occurs for high values of $\Delta \mu < \Delta \mu_{\text{HB}}$ where small regions synchronize for short periods of time, but system wide oscillations are not observed (\sfig{sfig:brussfield_preHopfTraj}). Furthermore, the transition moves closer to the macroscopic transition point with increased volume of the individual compartments (\sfig{sfig:brussfield_preHopfTraj}).

\tocless\section{Discussion}
    We have introduced a novel quantity, the entropy production factor $\epffield$, that quantifies irreversibility in macroscopic, non-equilibrium dynamics by measuring time-reversal symmetry breaking in the cross-covariances between multiple variables. Integrating $\epffield$ gives a lower bound on the net entropy production rate, $\eprfield$.

    Here, we illustrated that the total irreversibility rate cannot distinguish between the dynamical phase transitions in the well-mixed and the spatially extended Brusselator (\sfig{sfig:bruss_vs_brussfield_EPR}). While the EPR quantifies the emergence of oscillations, the synchronization of the oscillations across space is only captured in $\epffield$ by its peak shifting from high to low wavenumber (\fig{fig:epfPlot_brussfield}). By simulating systems with increasing compartment volumes, this shift occurs closer to the macroscopic transition point (\sfig{sfig:brussfield_preHopfTraj}), similarly to the increasing sharpness of the shift in $\epr$ for the well-mixed Brusselator (\fig{fig:bruss_eprScaling}). Thus, synchronization is intimately related to the emergence of oscillations. We hypothesize that synchronization occurs due the presence of a slow segment of the Brusselator dynamics (\fig{fig:epr_brusselator}a). The time spent in the slow portion of the dynamics allows neighboring oscillators to reduce their relative phase through their diffusive coupling, allowing previously out-of-sync lattice sites to synchronize via the low-cost mechanism of diffusion. This is further seen by the higher value of $\eprfield_\mathrm{blind}$ for the reaction-diffusion Brusselator compared to $\epr_\mathrm{blind}$ for the well-mixed Brusselator when $\Delta \mu < \Delta \mu_\mathrm{HB}$, but not for $\Delta \mu > \Delta \mu_\mathrm{HB}$ (\sfig{sfig:bruss_vs_brussfield_EPR}). Once the oscillations are synchronized, diffusion between lattice sites at equal concentrations is an equilibrium process and does not produce entropy.

    Previous work has investigated the behavior of $\epr$ at thermodynamic phase transitions of various kinds in an Ising model~\cite{Zhang2016}, but our work is only concerned with dynamical phase transitions. While~\cite{Nguyen2018} found $\epr$ to have a discontinuity of its first derivative with respect to $\Delta \mu$ in a slightly modified version of the well-mixed Brusselator, work on the same system presented here did not find any non-analytic behavior in $\epr_\mathrm{true}$~\cite{Rana2020}. We show that a discontinuous phase transition exists in our model, but the magnitude of the discontinuity is small and difficult to detect in $\epr_\mathrm{true}$ and is more easily seen in the coarse-grained $\epr_\mathrm{blind}$ (\fig{fig:bruss_eprScaling}). Other spectral decompositions of the dissipation rate either assume a particular form for the underlying dynamics~\cite{Tociu2019} or require the measurement of a response function in addition to the correlation function~\cite{Nardini2017}, which is often difficult in experiments.

    Calculating $\epffield$ does not require knowledge about the form of the underlying dynamics and is easy to calculate for many types of data, including both random variables, such as the positions of driven colloidal particles~\cite{Sun2009BrownianVortexes} (\sfig{sfig:eprPlot_dbp} \& \ref{sfig:dbp_234dim}), and random fields, such as spatially heterogeneous protein concentrations in cells~\cite{Bement2015}. Furthermore, we stress that we are only able to resolve the irreversibility present in the observable dynamics of our chemical example. As discussed above, the presence of hidden dynamics will provide underestimates of irreversibility measured via~\eq{eq:eprDef} due to the data processing inequality~\cite{Cover2006ElementsTheory}. Using other observable information, such as asymmetric transition rates~\cite{Martiniani2019QuantifyingEquilibrium} or the ratio of populations in observed states under stalled conditions~\cite{Polettini2017EffectiveObserver} in Markov jump processes, can give tighter bounds on the entropy produced when unobserved, dissipative processes are present. While the examples considered here are simulations of 1+1 dimensional fields, there is nothing inherently different in the methodology used here if one were to analyze experimental data in 2 or 3 spatial dimensions, such as the 3+1 dimensional time series data attained using lattice-light sheet microscopy~\cite{Chen2014LatticeResolution}.

    In active matter, both living and non-living, the non-equilibrium dissipation of energy manifests in both time and space. With the method introduced here, compatible with widely-used computational and experimental tools, we provide access to these underexplored modes of irreversibility that drive complex spatiotemporal dynamics.

\tocless\section{Acknowledgements}
    We would like to thank Samuel J. Bryant, Pranav Kantroo, Maria P. Kochugaeva, Rui Ma, Pierre Ronceray, A. Pasha Tabatabai, John D. Treado, Artur Wachtel, Dong Wang, and Vikrant Yadav for insightful discussions. DSS acknowledges support from NSF Fellowship  grant $\#$DGE1122492. BBM acknowledges a Simons Investigator award in MMLS. MPM and DSS acknowledge support from Yale University Startup Funds. MPM acknowledges funding from ARO MURI W911NF-14-1-0403, NIH RO1 GM126256, Human Frontiers Science Program (HFSP) grant $\#$ RGY0073/2018, and NIH U54 CA209992. Any opinion, findings, and conclusions or recommendations expressed in this material are those of the authors and do not necessarily reflect the views of the NSF, NIH, HFSP, or Simons Foundation.

\tocless\section{Methods}
    \tocless\subsection{Calculating $\epf$ from data}
        Estimate $\epf$ requires estimating frequency-space covariance functions, or cross spectral densities (CSDs). Considering a set of $M$ discrete, real variables measured over time: $\lbrace X^i (t) \rbrace$, where $t = \Delta t, \ldots, T$, with $T=N \Delta t$, and $i = 1, \ldots, M$ indexes the variables, we estimate the CSD using the periodogram,
        \begin{equation}
            \label{eq:periodogram}
            \raw{C}^{ij} (\omega_n) = \dfrac{1}{N^2} x^i(\omega_n) x^j(-\omega_n)
        \end{equation}
        \noindent where $x^i(\omega) = \mathcal{F} \lbrace X^i(t) - \mean{X^i(t)} \rbrace$ are the Fourier transforms of the centered variables over the frequencies $\omega_n = 2\pi n / T$ for $n=[-N/2, N/2]$.

        The periodogram, is known to exhibit a systematic bias and considerable variance in estimating the true CSD. Both of these issues can be resolved by smoothing $\raw{C}^{ij}$ via convolution with a Gaussian with width $\sigma$. This is equivalent to multiplying $\raw{C}^{ij}$ in the time domain by a Gaussian of width $1/\sigma$. We then define our smoothed CSD as
        \begin{equation}
            \label{eq:smoothPSD}
            \smooth{C}^{ij}(\omega_n) = \sum_{\omega_\mu} \Delta \omega \frac{\exp[-(\omega_\mu - \omega_n)^2/2 \sigma^2]}{\sqrt{2 \pi \sigma^2}} \raw{C}^{ij}(\omega_\mu)
        \end{equation}

        Once $\smooth{C}$ is calculated, we then use the discrete version of~\eq{eq:spectralEPR_particles} to estimate $\epf$. The extension to higher-dimensional data is done as follows: taking into account the spatial lattice on which the data is taken in~\eq{eq:periodogram}, convolving the result with a multivariate Gaussian in~\eq{eq:smoothPSD}, and finally estimate $\eprfield$ using the discrete version of~\eq{eq:spectralEPR_fields}. The choice of smoothing width, $\sigma$, should be guided by the maximum curvature seen in the structure factor, $C^{ij}$~\cite{Wang2016EntropySeparation}.

    \tocless\subsection{Bias in $\smooth{\epf}$ and $\smooth{\epr}$}
        Our estimates of $\smooth{\epf}$ and $\smooth{\epr}$ are biased. The bias is found by calculating the expected value of $\smooth{\epr}$ for a system in equilibrium. To do this, we assume that the true covariance function is $C^{ij} = \delta^{ij}$ and measurement noise plus finite sampling time and rate gives rise to Gaussian noise in both the real and complex parts of $\raw{C}^{ij}(\omega)$, obeying the symmetries required for $C^{ij}$ to be Hermitian. We only cite the results here and refer the reader to the Supplementary Material for a full derivation. The bias for random variables is
        \begin{align}
            \epf_{\text{bias}} &= \dfrac{M(M-1)}{2} \dfrac{\sqrt{\pi}}{T \sigma} \\
            \epr_{\text{bias}} &= \dfrac{M(M-1)}{2} \dfrac{\omega_{max}}{T \sigma \sqrt{\pi}},
        \end{align}
        where $M$ is the number of variables, $\omega_{max}$ is the maximum frequency available, $\sigma$ is the width of the Gaussian used to smooth $\raw{C}(\omega)$, and $T$ is the total time. The bias for random fields is
        \begin{align}
             \epffield_{\text{bias}} &= \left(\dfrac{M(M-1)}{2} + \dfrac{3M}{8} \right) \dfrac{\sqrt{\pi}}{T \sigma_\omega} \prod\limits_{i=1}^d \dfrac{\sqrt{\pi}}{L_i \sigma_{q_i}} \\
             \eprfield_{\text{bias}} &= \left( \dfrac{M(M-1)}{2} + \dfrac{3 M}{8} \right) \dfrac{\omega_{max}}{T \sigma_\omega \sqrt{\pi}} \prod\limits_{i=1}^d\dfrac{q_{i, max}}{L_i \sigma_{q_i} \sqrt{\pi}},
        \end{align}
        where $L_i$ is the length, $q_{i,max}$ is the maximum wavenumber, and $\sigma_{q_i}$ is the width of the Gaussian used to smooth $\raw{C}(\mathbf{q}, \omega)$ in the $i^{th}$ spatial dimension.

    \tocless\subsection{Simulations}
    To simulate the driven Gaussian fields,~\eq{eq:gaussfield}, we nondimensionalize the system of equations using a time scale $\tau = 1/(Dr)$ and length scale $\lambda = 1/\sqrt{r}$. We use an Euler-Maruyama algorithm to simulate the dynamics of the two fields on a periodic, 1 dimensional lattice.

    We simulate~\eq{eq:brusselator} using Gillespie's algorithm~\cite{Gillespie1977} to create a stochastic trajectory through the $(X,Y)$ phase plane with a well-mixed volume of $V = 100$. We calculate the true $\epr$ of any specific trajectory $\mathbf{z} = \lbrace m_j | j=1, \ldots, N \rbrace$ as follows. For each state $m'$, there exists a probability per unit time of transitioning to a new state $m$ via a chemical reaction $\mu$, denoted by $W_{m , m'}^{(\mu)}$. At steady state, the true entropy produced is~\cite{Esposito2010ThreeTheorems}
    \begin{equation}
        \label{eq:markovEP}
        \Delta S_{\text{true}} [\mathbf{z}] = \sum_{j = 1}^N \ln \dfrac {W_{m_j , m_{j - 1}}^{(\mu_j)}}{W_{m_{j - 1}, m_j}^{(\mu_j)}}
    \end{equation}
    Note that $\Delta S_{\text{true}}$ is now itself a random variable that depends on the specific trajectory. We estimate $\mean{\epr_{\text{true}}}$ by fitting a line to an ensemble average of $\Delta S_{\text{true}}$ (\sfig{sfig:brussFit}), and compare that to $\smooth{\epr}$. We calculate $\epr_{\text{blind}}$ by considering the ``rate" at which a transition can occur as the sum over all the rates that give rise to the observed transition in $(X, Y)$, i.e.
    \begin{equation}
        \Delta S_{\text{blind}}=\sum_{j=1}^{N} \ln  \dfrac{\sum\limits_{\lbrace \mu_j | m_{j-1} \rightarrow m_j \rbrace} W_{m_j, m_{j-1}}^{( \mu_j)}}{\sum\limits_{\lbrace \mu_j | m_{j-1} \rightarrow m_j \rbrace} W_{m_{j-1}, m_j}^{(\mu_j)}}
    \end{equation}

    To simulate the reaction-diffusion Brusselator,~\eq{eq:brussfield}, we take a compartment-based approach~\cite{Erban2007AProcesses} where we treat each chemical species in each compartment as a separate species, and treat diffusion events as additional chemical reaction pathways. We nondimensionalize time by $\tau = 1/k_1^+$ and use a Gillespie algorithm to simulate all reactions on a 1 dimensional periodic lattice with $L$ sites.

    See Supplementary Materials for all simulation parameters used in each figure.

\tocless\section{Data Availability}
The data that support the findings of this study are available from the corresponding authors upon reasonable request.

\tocless\section{Code Availability}
The code used to calculate the EPR and EPF from data, as well as run all the simulations in this study, can be found at \href{https://github.com/lab-of-living-matter/freqent}{this Github page}.

\bibliography{refs.bib}

\appendix

\pagebreak
\newgeometry{margin=0.75in}
\begin{center} 
    \large{\textbf{Dissipative signatures of dynamical phases and transitions \\ Supplementary Materials}}
\end{center}

\setcounter{figure}{0}
\renewcommand{\figurename}{Supplementary Figure}
\renewcommand{\baselinestretch}{0.5}

\tableofcontents

\section{Derivation of the entropy production factor}
    Consider a system described by a set of real random scalar variables tracing some path through phase space, $\mathbf{X} = \lbrace X^i(t) \rbrace$ with Fourier transforms given by $x^i(\omega)$. Assuming the variables to be Gaussian distributed with real space covariance function with time translation invariance, $\mean{ X^i(t) X^j(t') } = C^{ij}(t-t')$, the probability of observing a particular path is given in frequency space as
    \begin{equation}\label{eq:gaussPathProb}
        \mathcal{P} \left[\mathbf{x}\right] = \dfrac{1}{Z} \exp \left[ -\dfrac{1}{2}\int\limits_{-\infty}^{\infty} \dfrac{d \omega}{2 \pi} C^{-1}_{ij}(\omega) x^i(-\omega) x^j(\omega) \right]
    \end{equation}
    where $Z$ is the partition function and $C^{ij}(\omega) = \mathcal{F} \left[ C^{ij}(t-t') \right]$ with $C_{ij}^{-1} \equiv (C^{-1})_{ij}$. The reverse path is given by \eq{eq:gaussPathProb} with $\omega \to -\omega$ in the argument of $C^{-1}_{ij}$. 
    \begin{equation}
        \label{eq:gaussPathProbReverse}
        \widetilde{\mathcal{P}}\left[ \mathbf{x} \right] = \dfrac{1}{\widetilde{Z}} \exp \left[ -\dfrac{1}{2}\int\limits_{-\infty}^{\infty} \dfrac{d \omega}{2 \pi} C^{-1}_{ij}(-\omega) x^i(-\omega) x^j(\omega) \right].
    \end{equation}

    To make the following calculations easier, we consider a discrete case for a finite time series of length $T = N dt$ with sampling rate $dt$. In this case, \eq{eq:gaussPathProb} is written as
    \begin{equation}
        \label{eq:discreteGaussPathProb}
         \mathcal{P} \left[ \mathbf{x} \right] = \dfrac{1}{Z} \prod\limits_{n=1}^{T/dt} \exp \left[ -\dfrac{1}{2T} \left[ C^{-1}(\omega_n) \right]_{ij} x^i(-\omega_n) x^j(\omega_n) \right]
    \end{equation}
    where $\omega_n = 2\pi n/T$. \eq{eq:gaussPathProbReverse}
    is written similarly. We then have
    \begin{equation}
        \label{eq:logProbs}
        \ln \left( \dfrac{\mathcal{P}}{\widetilde{\mathcal{P}}} \right) = \ln \left( \dfrac{\widetilde{Z}}{Z} \right) + \dfrac{1}{2T}\sum\limits_{n=1}^{T/dt} \left[ C^{-1} (-\omega_n)  - C^{-1} (\omega_n) \right]_{ij} x^i(-\omega_n) x^j(\omega_n)
    \end{equation}
    Using the fact that, for a finite signal of length $T$, $\mean{ x_i(\omega_n) x_j(-\omega_m) } = T \delta_{nm} C_{ij}(\omega_n)$, the KL-divergence is then
    \begin{align}
        D_{KL} \left( \mathcal{P} \left[\mathbf{x}\right] \ \middle \| \ \widetilde{\mathcal{P}}\left[ \mathbf{x} \right] \right) &= \mean{ \ln \left( \dfrac{\mathcal{P}}{\widetilde{\mathcal{P}}} \right)}_{\mathcal{P}}\\
         &= \sum_\mathbf{x} \, \mathcal{P} \ln \left( \dfrac{\mathcal{P}}{\widetilde{\mathcal{P}}} \right) \\
         &= \ln \left( \dfrac{\widetilde{Z}}{Z} \right) + \dfrac{1}{2} \sum\limits_n \left[ C^{-1} (-\omega_n)  - C^{-1} (\omega_n) \right]_{ij} C^{ji}(\omega_n)
    \end{align}
    where $\sum_\mathbf{x}$ is a sum over all possible paths. The entropy production rate is then given by
    \begin{equation}
        \label{eq:spectralEPR_particles_discrete}
        \epr = \lim_{T \to \infty} \dfrac{1}{2T} \sum\limits_n \left[ C^{-1} (-\omega_n)  - C^{-1} (\omega_n) \right]_{ij} C^{ji}(\omega_n) = \lim_{T \to \infty} \dfrac{1}{T} \sum\limits_n \epf(\omega_n),
    \end{equation}
    where we have introduced the entropy production factor, $\epf$, and dropped the ratio of the partition functions as they will contribute $0$ to the EPR when multiplied by $1/T$ and taking the $T \to \infty$ limit. In the limit taken, the sum becomes an integral, $\sum_n \to T/2\pi \int d \omega$, which brings us to our first main result
    \begin{equation}
        \label{seq:spectralEPR_particles}
        \epr = \int\limits_{-\infty}^{\infty} \dfrac{d \omega}{2 \pi} \ \epf(\omega)
    \end{equation}
    $\epf$ can also be rewritten as a $\epf(\omega) = \mathrm{Tr} \lbrace \mathbf{C}(\omega) \left[ \mathbf{C}^{-1}(-\omega) - \mathbf{C}^{-1}(\omega) \right] \rbrace / 2$.
    
    In the same spirit as above, we derive a similar expression for a set of real-valued random fields, $\boldsymbol{\eta}(\mathbf{r},t) = \{ \phi^i (\mathbf{r},t) | \mathbf{r} \in \mathbb{R}^d \}$. The probability of following a path is given by
    \begin{equation}
        \label{eq:pathProbRF}
        \mathcal{P} [\boldsymbol{\eta}] = \dfrac{1}{Z} \exp \left[-\dfrac{1}{2} \iint \dfrac{d \omega}{2 \pi} \dfrac{d^d \mathbf{q}}{(2 \pi)^d} C^{-1}_{ij}(\mathbf{q}, \omega) \phi^i(-\mathbf{q}, -\omega) \phi^j(\mathbf{q}, \omega) \right],
    \end{equation}
    where the covariance function is defined as $\mean{ \phi^i(\mathbf{q}, \omega) \phi^j(\mathbf{q}', \omega') } = C^{ij}(\mathbf{q}, \omega) \delta^d(\mathbf{q} + \mathbf{q}') \  \delta(\omega + \omega')$. Under time reversal, $\mathbf{q}$ is invariant but $\omega$ switches sign. Thus, $\mathcal{P} [\widetilde{\boldsymbol{\eta}}]$ is given by
    \begin{equation}
        \label{eq:pathProbRF_reverse}
        \mathcal{P} [\widetilde{\boldsymbol{\eta}}] = \dfrac{1}{Z} \exp \left[ -\dfrac{1}{2} \iint \dfrac{d \omega}{2 \pi} \dfrac{d^d \mathbf{q}}{(2 \pi)^d} C^{-1}_{ij}(\mathbf{q}, -\omega) \phi^i(-\mathbf{q}, -\omega) \phi^j(\mathbf{q}, \omega) \right].
    \end{equation}

    We assume that the field is sampled in time with resolution $dt$ for a time $T$ (i.e. $\Delta \omega = 2 \pi /T$) and each dimension of space is sampled with resolution $dx_i$ for a length $L_i$ (i.e. $\Delta k_i = 2 \pi /L_i$), giving the discretized path probability functional
    \begin{equation}
        \mathcal{P}[\boldsymbol{\eta}] = \dfrac{1}{Z} \prod_{n=0}^{T/dt} \prod_{m_1=0}^{L_1/dx_1} \ldots \prod_{m_d=0}^{L_d/dx_d} \exp \left[ -\dfrac{1}{2TV}  \ C^{-1}_{ij}(\mathbf{q}_m, \omega_n) \phi^i(-\mathbf{q}_m, -\omega_n) \phi^j(\mathbf{q}_m, \omega_n) \right],
    \end{equation}
    where $\mathbf{q}_m \equiv (k_{1_{m_1}}, k_{2_{m_2}}, \ldots, k_{d_{m_d}})$ and $V$ is the total volume. A similar expression exists for $\mathcal{P} [\widetilde{\boldsymbol{\eta}}]$. From these expressions, we have (ignoring the partition functions that will add zero once we take the $T \to \infty$ limit)
    \begin{equation}
        \ln\left( \dfrac{\mathcal{P}}{\widetilde{\mathcal{P}}} \right) = \dfrac{1}{2VT} \sum_{n=0}^{T/dt} \sum_{m_1=0}^{L_1/dx_1} \ldots \sum_{m_d=0}^{L_d/dx_d} \Big[ C^{-1}_{ij}(\mathbf{q}_m,-\omega_n) - C^{-1}_{ij}(\mathbf{q}_m, \omega_n) \Big] \phi^i(-\mathbf{q}_m, -\omega_n) \phi^j(\mathbf{q}_m, \omega_n).
    \end{equation}
    
    Taking the average with respect to $\mathcal{P}$ and noting that \\ $\mean{ \phi^i(\mathbf{q}_m, \omega_n) \phi^j(-\mathbf{q}_{m'}, -\omega_{n'})} = T V \delta_{m m'} \delta_{n n'} C^{ij}(\mathbf{q}_m, \omega_n)$ for finite signals, we have
    \begin{equation}
        D_{KL} \left( \mathcal{P}[\boldsymbol{\psi}] \ \middle \| \ \mathcal{P} [\widetilde{\boldsymbol{\eta}}] \right) = \dfrac{1}{2} \sum_{n=0}^{T/dt} \sum_{m_1=0}^{L_1/dx_1} \ldots \sum_{m_d=0}^{L_d/dx_d} \left[ C^{-1}(\mathbf{q}_m,-\omega_n) - C^{-1}(\mathbf{q}_m, \omega_n) \right]_{ij} C^{ji}(\mathbf{q}_m, \omega_n).
    \end{equation}
    This gives an entropy production rate of
    \begin{equation}
        \label{eq:spectralEPR_fields_discrete}
        \epr = \lim_{T\to\infty} \dfrac{1}{2T} \sum_{n=0}^{T/dt} \sum_{m_1=0}^{L_1/dx_1} \ldots \sum_{m_d=0}^{L_d/dx_d} \epffield(\mathbf{q}_m, \omega_n),
    \end{equation}
    \noindent again introducing the EPF for fields, $\epf(\mathbf{q}, \omega)$. Passing to the continuum limit, we have
    \begin{equation}
        \label{seq:spectralEPR_fields}
        \epr = V \int\limits_{-\infty}^{\infty} \dfrac{d \omega}{2\pi} \dfrac{d^d q}{(2 \pi)^d} \epf(\mathbf{q}, \omega),
    \end{equation}
    The EPR density is given by $\eprfield = \epr / V$. As with the case of random variables, $\epffield$ can be rewritten as a trace, $\epffield(\mathbf{q}, \omega) = \mathrm{Tr} \lbrace \left[ \mathbf{C}^{-1}(\mathbf{q}, -\omega) - C^{-1}(\mathbf{q}, \omega) \right] \mathbf{C}(\mathbf{q}, \omega) \rbrace / 2$.
    
    \subsection{Proof of lower bound}
        The KL divergence in Eq. 1 is an exact expression for the entropy production rate provided that the observed set of variables, $\lbrace x^\mu \rbrace$, contains every non-equilibrium degree of freedom present in the system. In practice, one only has access to a subset of those degrees of freedom, making the measured KL divergence a lower bound on the entropy production rate. Here, we show that the Gaussian assumption for P[X] provides another lower bound on the irreversibility measured on the scale of the observed mesoscale trajectories.

The proof relies on the data processing inequality~\cite{Cover2006ElementsTheory}, which states that any transformation of variables $F: x^\mu \to y^\mu$ will lower the relative entropy between two distributions over both sets of variables, i.e. 
        \begin{equation}
            D_{KL} \left( P \left[ \lbrace x^\mu \rbrace \right] \vert \vert Q \left[ \lbrace x^\mu \rbrace \right] \right) \geq D_{KL} \left( P \left[ \lbrace y^\mu \rbrace \right] \vert \vert Q \left[ \lbrace y^\mu \rbrace \right] \right).
        \end{equation}
        Intuitively, it states that any processing of an observation $\lbrace x^\mu \rbrace$ makes it more difficult to determine whether it came from $P$ or $Q$. Our strategy will be to choose a transformation that will turn any distribution over $x^\mu$ into a Gaussian distribution over $y^\mu$. In our case, our observables are the frequency space variables $x^\mu (\mathbf{q}, \omega)$, and the transformation is a multiplication of by a random phase field $\theta(\mathbf{q}, \omega)$, i.e. $x^\mu (\mathbf{q}, \omega) \to x^\mu(\mathbf{q}, \omega) e^{i \theta(\mathbf{q}, \omega)}$. This random phase, when integrated over frequency space, will make all correlations zero except for the two-point correlation function due to the fact that the variables in real space are real, making the two-point correlation equal to $\langle x^\mu (x^\mu)^* \rangle$, cancelling the random phase. Thus, the transformed variables are described by a Gaussian distribution (defined as the distribution whose only non-zero cumulants are the first and second), and the data processing inequality guarantees that this provides a lower bound to the KL divergence over the original distributions.

    \subsection{Numerical calculations}
        \subsubsection{Partition functions}
            While we were able to drop the partition functions in the analytic calculations above after taking the $T \to \infty$ limit and assuming translation and rotation invariance, the case of finite, discrete data requires greater care due to the presence of noise. The partition function for \eq{eq:pathProbRF} is given by
            \begin{equation}
                \label{eq:partition_RF}
                Z = \exp \left(  \dfrac{VT}{2} \iint \dfrac{d \omega}{2\pi} \dfrac{d \mathbf{q}}{2 \pi} \ \ln \left[ \det C (\mathbf{q}, \omega) \right]  \right).
            \end{equation}
            and the EPR density is given by
            \begin{equation}
                \label{eq:spectralEPR_fields_partition}
                \eprfield =\frac{1}{2} \iint \frac{d^{d} \mathbf{q}}{(2 \pi)^{d}} \frac{d \omega}{2 \pi} \left[ \ln \left(\frac{\det C(\mathbf{q}, -\omega)}{\det C(\mathbf{q}, \omega)} \right) + \left[ C^{-1}(\mathbf{q}, -\omega) - C^{-1} (\mathbf{q}, \omega) \right]_{i j} C^{ji}(\mathbf{q}, \omega) \right].
            \end{equation}
            In discrete form,
            \begin{equation}
                \label{eq:spectralEPR_fields_partition_discrete}
                \eprfield = \frac{1}{2TV} \sum_{m,n}  \left[ \ln \left( \dfrac{\det C(\mathbf{q}_m, -\omega_n)}{\det C(\mathbf{q}_m, \omega_n)} \right) + \left[ C^{-1}(\mathbf{q}_m, -\omega_n) - C^{-1} (\mathbf{q}_m, \omega_n) \right]_{i j} C^{ji}(\mathbf{q}_m, \omega_n) \right]
            \end{equation}
            
            For random variables, the equivalent of \eq{eq:partition_RF} has an integral over only $\omega$ in the argument of the exponential resulting in the same value of $\tilde{Z}$ since the integral is over all $\omega$. This gives $\ln(\widetilde{Z} / Z) = 0$, allowing one to drop the partition functions even for finite trajectories. However, for finite trajectories of fields, the partition functions are required to maintain the positivity of the relative entropy used to define $\eprfield$.
    
        \subsubsection{First moment properties}
            We now turn to the problem of estimating the bias in our measured entropy production rate. For this, we assume that we have an equilibrium process and calculate what the average measured entropy production rate is, representing the systematic overestimation of our estimator. We work in coordinates where the covariance matrix is the identity, $C^{\mu \nu} = \delta^{\mu \nu}$. Due to a combination of measurement errors and only having a finite time series, we will measure a matrix that deviates from the identity by
            \begin{equation}
                \raw{C}^{\mu \nu} = \delta^{\mu \nu} + \raw{R}^{\mu \nu} + i \raw{A}^{\mu \nu},
            \end{equation}
            where $\raw{R}^{\mu \nu}(\omega)$ and $\raw{A}^{\mu \nu}(\omega)$ are elements of a symmetric and anti-symmetric $D \times D$ matrix, respectively, each assumed to be much smaller than 1. The anti-symmetric contribution must be purely imaginary because $\raw{C}^{\mu \nu}$ is a Hermitian matrix by definition. Further, we have $R^{\mu \nu}(-\omega) = R^{\mu \nu}(\omega)$ and $A^{\mu \nu}(-\omega) = -A^{\mu \nu}(\omega)$. For notational simplicity, we define $\raw{M}^{\mu \nu} \equiv \delta^{\mu \nu} + \raw{R}^{\mu \nu}$ and therefore $\mathbf{\smooth{C}} = \mathbf{\smooth{M}} + i \mathbf{\smooth{A}}$.
            
            To calculate the EPR, we need to calculate the EPF, $\epf = \mathrm{Tr} \lbrace \mathbf{C}(\omega) \left[ \mathbf{C}^{-1}(-\omega) - \mathbf{C}^{-1}(\omega) \right] \rbrace$. We approximate $\mathbf{\smooth{C}}^{-1}$ as 
            \begin{equation}
                \label{eq:cInv_approx}
                \mathbf{\smooth{C}}^{-1} = (\mathbf{\smooth{M}} + i \mathbf{\smooth{A}})^{-1} \approx \mathbf{\smooth{M}}^{-1} - i \mathbf{\smooth{M}}^{-1} \mathbf{\smooth{A}} \mathbf{\smooth{M}}^{-1}.
            \end{equation}
            Then, $\mathbf{C}^{-1}(-\omega) - \mathbf{C}^{-1}(\omega) = 2i \mathbf{\smooth{M}}^{-1} \mathbf{\smooth{A}} \mathbf{\smooth{M}}^{-1}$. Multiplying by $\mathbf{\smooth{C}}$,
            \begin{equation}
                \label{eq:CCasym}
                \mathbf{C}(\omega) \left[ \mathbf{C}^{-1}(-\omega) - \mathbf{C}^{-1}(\omega) \right] \rbrace = 2i \mathbf{\smooth{M}}^{-1} \mathbf{\smooth{A}} \mathbf{\smooth{M}}^{-1} \mathbf{\smooth{M}} - 2 \mathbf{\smooth{M}}^{-1} \mathbf{\smooth{A}} \mathbf{\smooth{M}}^{-1} \mathbf{\smooth{A}} 
            \end{equation}
            Taking the trace of \eq{eq:CCasym}, the first term is an asymmetric matrix with zero trace. By writing $\mathbf{\smooth{M}} = \mathbb{I} + \mathbf{\smooth{R}}$, we approximate $\mathbf{\smooth{M}}^{-1} \approx \mathbb{I} - \mathbf{\smooth{R}}$, and the second term is approximately as $\mathbf{\smooth{A}}^2 + \mathcal{O}(\mathbf{\smooth{A}}^2 \mathbf{\smooth{R}} + \mathbf{\smooth{A}} \mathbf{\smooth{R}} \mathbf{\smooth{A}})$. Thus, to lowest order we have
            \begin{equation}
                \epf = \mathrm{Tr} \lbrace \mathbf{C}(\omega) \left[ \mathbf{C}^{-1}(-\omega) - \mathbf{C}^{-1}(\omega) \right] \rbrace \approx -  \mathrm{Tr} \left( \mathbf{\smooth{A}}^2 \right) = 2 \sum_{\mu > \nu} (\smooth{A}^{\mu \nu})^2.
            \end{equation}
            Assuming each element of $\mathbf{A}$ is an iid random variable, we can write the average EPF measured at equilibrium as
            \begin{equation}
                \mean{\epf}_{eq} = 2 \dfrac{M(M-1)}{2} \mean{(\smooth{A}^{\mu \nu})^2} = \dfrac{M(M-1)}{2}.
            \end{equation}
            We calculated $\mean{ (\raw{A}^{\mu \nu})^2}$ as follows:
            \begin{align}
                \mean{(\raw{A}^{\mu \nu})^2} &= \mean{\left[ \mathrm{Im}(x^\mu x^{\nu*}) \right]^2} \\
                &= \mean{ \left[ \mathrm{Re}(x^\mu) \mathrm{Im}(x^{\nu*}) + \mathrm{Im}(x^\mu) \mathrm{Re}(x^{\nu*})\right]^2 } \\
                &= \mean{\left[\mathrm{Re}(x^\mu) \mathrm{Im}(x^{\nu*})\right]^2 + \left[\mathrm{Im}(x^\mu) \mathrm{Re}(x^{\nu*})\right]^2 + \text{cross-terms}} \\
                &= 1/2 \label{eq:ARawSquared_mean}
            \end{align}
            The cross-terms average to zero because, in our choice of coordinate system, $|x^\mu|^2 = 1$, so the real and imaginary parts of $x^\mu$ are equally distributed along the unit circle. In addition, $x^\mu$ and $x^\nu$ should be uncorrelated (recall that we are working at equilibrium). The first and second term each have both a real and an imaginary part squared, each of which is always positive and on average equal to $1/2$. Thus, each term is $1/4$ and adds to $1/2$.
            
            To estimate $\smooth{\epr}$, we smooth $\raw{\epf}$ and integrate over all frequencies. We calculate $\mean{ (\smooth{A}^{\mu \nu})^2 }$ as
            \begin{align}
                (\smooth{A}^{\mu \nu})^2 &= \left( \sum_{\omega_i} \Delta \omega \frac{\exp[-(\omega_i - \omega_n)^2/2 \sigma^2]}{\sqrt{2 \pi \sigma^2}} \raw{A}_{\mu \nu}(\omega_i) \right)^2 \\
                \mean{ (\smooth{A}^{\mu \nu})^2 } &= \sum_{\omega_i} (\Delta \omega)^2 \frac{\exp[-(\omega_i - \omega_n)^2/\sigma^2]}{2 \pi \sigma^2} \mean{ (\smooth{A}^{\mu \nu})^2 (\omega_i) } \\
                &\approx \dfrac{\Delta \omega}{4 \pi \sigma^2} \int d \omega \exp[-\omega^2 / \sigma^2] = \dfrac{\Delta \omega}{4 \pi \sigma^2} \sqrt{\pi \sigma^2} \\
                &= \dfrac{\sqrt{\pi}}{2 T \sigma}.
            \end{align}
            We used $\mean{ \raw{A}^{\mu \nu}(\omega_i) \raw{A}^{\mu \nu}(\omega_j) } = \mean{(\raw{A}^{\mu \nu})^2(\omega_i) } \delta_{ij}$ in the second line, passed to an integral using one of the integration measures $\Delta \omega$ in the third line, and substituted $\Delta \omega = 2 \pi / T$ in the fourth line. Finally, we arrive at
            
            \begin{equation}
                \epr_{eq} = 2 \dfrac{M(M-1)}{2} \int\limits_{-\omega_{max}}^{\omega_{max}} \dfrac{d \omega}{2\pi} \mean{ (\smooth{A}^{\mu \nu})^2 } = \dfrac{M(M-1)}{2} \dfrac{\omega_{max}}{T \sigma \sqrt{\pi}}
            \end{equation}
            
            \noindent If we also average the covariance functions over $N$ independent trajectories with the same dynamics, this bias is further reduced, leaving us with our final estimate of the bias in our entropy production rate estimator
            \begin{equation}
                \label{eq:eprBias}
                \epr_{eq} = \dfrac{1}{N} \dfrac{M(M-1)}{2} \dfrac{\omega_{max}/\sigma}{T \sqrt{\pi}}
            \end{equation}
            
            Following the same line of reasoning for a set of $M$ fields in $d+1$ dimensions, a similar expression can be derived. We again write $C^{\mu \nu}(\mathbf{q}, \omega) = \mathbb{I} + R^{\mu \nu}(\mathbf{q}, \omega) + i A^{\mu \nu}(\mathbf{q}, \omega)$.  Extra care must be taken in the field case because \eq{seq:spectralEPR_fields} does not have the same symmetries as \eq{seq:spectralEPR_particles}. Specifically, while $R^{\mu \nu}(-\mathbf{q}, -\omega) = R^{\mu \nu} (\mathbf{q}, \omega)$ and $A^{\mu \nu}(-\mathbf{q}, -\omega) = -A^{\mu \nu} (-\mathbf{q}, -\omega)$, nothing can be said \textit{a priori} about $R(\mathbf{q}, -\omega)$ or $A(\mathbf{q}, -\omega)$.
            
            In order to calculate the bias, we will calculate the mean of the spatiotemporal entropy production factor, $\epffield =  \mathrm{Tr} \lbrace \left[ \mathbf{C}^{-1}(\mathbf{q}, -\omega) - \mathbf{C}^{-1}(\mathbf{q}, \omega) \right] \mathbf{C}(\mathbf{q}, \omega) \rbrace$. The calculation is tedious, so we only report the result here:
            \begin{align}
                \mean{ \epffield }_{eq} &= \mean{ \mathrm{Tr} \left[ \mathbf{R}^2 - \mathbf{A}^2 \right] } \\
                &= M(M-1) \left( \mean{ (R^{\mu \nu})^2 } + \mean{ (A^{\mu \nu})^2 } \right) + M \mean{ (R^{\mu \mu})^2 } \\
                &= M(M-1) + 3M/4
            \end{align}
            We arrived at this by using the fact that $\mathrm{Tr} \left(\mathbf{A}^2 \right) = \sum_{\mu \neq \nu}(A^{\mu \nu})^2$ and $\mathrm{Tr} \left(\mathbf{R}^2 \right) = \sum_{\mu} (R^{\mu \mu})^2 + \sum_{\mu \neq \nu} (R^{\mu \nu})^2$ for an asymmetric and symmetric matrix, respectively, in addition to the assumption that every matrix element is an i.i.d. random variable. As before, $\mean{ (A^{\mu \nu})^2 } = 1/2 = \mean{ (R^{\mu \nu})^2 }$. Now turning to the diagonal elements of $\mathbf{R}$,
            \begin{align}
                \mean{ (R^{\mu \mu})^2 } &= \mean{ \left[1 - \mathrm{Re} \left( \phi^\mu \phi^{\mu*} \right) \right]^2 } \\
                &= 1 + \mean{ \mathrm{Re} \left(\phi^\mu \right)^2 \mathrm{Im} \left( \phi^\mu \right)^2 } + \mean{ \mathrm{Re} \left( \phi^\mu \right)^4 } + \mean{ \mathrm{Im} \left( \phi^\mu \right)^4 } - 2 \mean{ \mathrm{Re} \left( \phi^\mu \right) ^2 + \mathrm{Im} \left( \phi^\mu \right) ^2 }  \\
                &= 3/4,
            \end{align}
            where we used $\mean{ x^4 } = 3 \mean{ x^2 } ^2$ for Gaussian variables and $\mean{ \mathrm{Re} \left( \phi^\mu \right)^2 } = \mean{ \mathrm{Im} \left( \phi^\mu \right)^2 } = 1/2$ in our choice of coordinate system.
            
            Assuming the signals to have a total length in time of $T$ and a total length in each spatial dimension of $L_i$, we smooth the spatiotemporal covariance function with a multivariate Gaussian of width $\sigma_{\omega}$ in the temporal dimension, and $\sigma_{k_i}$ in each of the spatial dimensions, giving a factor of $\omega_{max} / \sigma_\omega T \sqrt{\pi}$ for the temporal dimension and $q_{i, max} / \sigma_{q_i} L_i \sqrt{\pi}$ for each spatial dimension. Putting all these results together, we have
            \begin{equation}
                \label{seq:eprBiasField}
                \eprfield_{eq} =\dfrac{1}{N} \left( \dfrac{M(M-1)}{2} + \dfrac{3 M}{8} \right) \dfrac{\omega_{max}}{T \sigma_\omega \sqrt{\pi}} \prod\limits_{i=1}^d\dfrac{q_{i, max}}{L_i \sigma_{q_i} \sqrt{\pi}}
            \end{equation}

\section{EPR of coupled Gaussian fields}
    Consider the coupled equations of motion for the scalar fields $\phi$ and $\psi$ in $d+1$ dimensions.
    \begin{align}
        \label{seq:gaussfield_eom}
        \partial_t \phi(\mathbf{x}, t) &= -D (r - \nabla^2) \phi - \alpha \psi + \sqrt{2D} \xi_\phi \\
        \partial_t \psi(\mathbf{x}, t) &= -D (r - \nabla^2) \psi + \alpha \phi + \sqrt{2D} \xi_\psi.
    \end{align}
    \noindent with $\mean{ \xi^i(\mathbf{x}, t) \xi^j(\mathbf{x}', t') } = \delta^{ij} \delta(t-t') \delta^d(\mathbf{x}-\mathbf{x}')$. This is a Gaussian model with free energy
    \begin{equation}
        F = \int d^dx \ \left[ \dfrac{r}{2} \left( \phi^2 + \psi^2 \right) + \dfrac{1}{2} \left( |\nabla \phi|^2 + |\nabla \psi|^2 \right) \right].
    \end{equation}
    The interaction term cannot be written as a gradient of an energy, so we have
    \begin{align}
        \partial_t \phi(\mathbf{x}, t) &= -D \dfrac{\delta F}{\delta \phi} - \alpha \psi + \sqrt{2D} \xi_\phi \\
        \partial_t \psi(\mathbf{x}, t) &= -D \dfrac{\delta F}{\delta \psi} + \alpha \phi + \sqrt{2D} \xi_\psi.
    \end{align}
    Combining the two fields into a single vector, $\boldsymbol{\eta}(\mathbf{x}, t) = (\phi(\mathbf{x}, t), \psi(\mathbf{x}, t))^T$, we write: 
    \begin{equation}
        \label{seq:dgf_langevin}
        \partial_t \boldsymbol{\eta} = B \boldsymbol{\eta} + \sqrt{2D} \boldsymbol{\xi}; \qquad 
        B(\mathbf{x}) =
        \begin{pmatrix}
        -D (r - \nabla^2) & -\alpha \\
        \alpha & -D (r - \nabla^2)
        \end{pmatrix}.
    \end{equation}
    
    To get the cross-spectral density, we rewrite \eq{seq:dgf_langevin} as an Ito stochastic differential equation:
    \begin{equation}
        \label{eq:spinOscSDE}
        d\boldsymbol{\eta} = B \boldsymbol{\eta} \, dt + \Xi \,  d\mathbf{W},
    \end{equation}
    where $\mathbf{W}(\mathbf{x}, t)$ is a multidimensional Wiener process in space and time with strength $\Xi^{ij} = \sqrt{2D} \delta^{ij}$. The eigenvalues of $A$ have negative real parts, so a stationary solution exists. The cross-spectral density is~\cite{Gardiner2010StochasticSciences}
    \begin{equation}
        C(\mathbf{q}, \omega) = \left( B(\mathbf{q}) - i\omega \mathbb{I} \right)^{-1} \Xi \Xi^T  \left( B(\mathbf{q}) + i\omega \mathbb{I} \right)^{-T}.
    \end{equation}
    Noting $\mathbf{\Xi} \ \mathbf{\Xi}^T = 2D \mathbb{I}$, we have
    \begin{equation}
    \label{eq:corrMatrix_freqSpace}
        C(\omega) = \dfrac{2D}{|\left( D(r + q^2)  + i\omega \right)^2 + \alpha^2)|^2}
        \begin{pmatrix}
            \left[ D(r + q^2) \right]^2 + \alpha^2 + \omega^2 & i 2 \alpha \omega \\
            -i 2 \alpha \omega & \left[ D(r + q^2) \right]^2 + \alpha^2 + \omega^2
        \end{pmatrix}.
    \end{equation}
    The inverse is given by
    \begin{equation}
        C^{-1}(\omega) = \dfrac{1}{2D}
        \begin{pmatrix}
            \left[ D(r + q^2) \right]^2 + \alpha^2 + \omega^2 & -i 2 \alpha \omega \\
            i 2 \alpha \omega & \left[ D(r + q^2) \right]^2 + \alpha^2 + \omega^2
        \end{pmatrix}.
    \end{equation}
    Finally, using \eq{seq:spectralEPR_fields}, we have
    \begin{equation}
        \label{seq:epf_dgf}
        \epr = V \iint\limits_{-\infty}^{\infty} \dfrac{d \omega}{2 \pi}\dfrac{d \mathbf{q}}{2 \pi} \dfrac{8 \alpha^2 \omega^2}{|( D(r + q^2)  + i\omega)^2 + \alpha^2)|^2} = V \dfrac{\alpha^2}{D \sqrt{r}}.
    \end{equation}
    Rearranging the denominator of the integrand of above gives $\epffield^{DGF}$ given in the main text.
    
    We can alternatively calculate the entropy production rate by using the Onsager-Machlup functional~\cite{Onsager1953} for the path probability functional $P[\boldsymbol{\eta}]$ in
    \begin{equation}
        \epr = \lim_{T\to\infty} \dfrac{1}{T} \mean{ \ln \dfrac{P[\boldsymbol{\eta}]}{\widetilde{P}[\boldsymbol{\eta}]} }
    \end{equation}
    Writing it as a path $P[\boldsymbol{\eta}] \propto \exp(-\mathcal{A})$, where $\mathcal{A}$ is the action, this becomes
    \begin{equation}
        \epr = \lim_{T\to\infty} \dfrac{1}{T} \mean{ \widetilde{\mathcal{A}} - \mathcal{A} },
    \end{equation}
    where $\widetilde{\mathcal{A}}$ is the action under time-reversal. To calculate $\mathcal{A}$, we use standard path integral techniques, i.e. the Martin-Siggia-Rose formalism~\cite{Martin1973StatisticalSystems}. The idea is to try and find the expectation of some observable, $O$, over noise realizations.
    \begin{equation}
        \mean{ O[\boldsymbol{\eta}] }_\xi = \int \mathcal{D}[\boldsymbol{\xi}] \ O[\boldsymbol{\eta}] P[\boldsymbol{\xi}].
    \end{equation}
    Since the noise is Gaussian, we have
    \begin{equation}
        P[\boldsymbol{\xi}] \propto \exp \left( \dfrac{1}{4D} \int d^dx \ dt \ \boldsymbol{\xi}^2 \right)
    \end{equation}
    (we use Einstein notation throughout). We then insert the most complicated expression for 1 ever written. Using the integral representation of the functional delta function, $\delta[f(x)] = \int \mathcal{D}[i\widetilde{f}] \exp[-\int dx \widetilde{f}(x) f(x)]$, we write
    \begin{align}
        1 &= \int \prod_j \mathcal{D} [\eta^j] \delta(\partial_t \eta^j - B^j_k \eta^k - \xi^j) \\
        &= \int \prod_j \mathcal{D} [\eta^j] \mathcal{D}[i\widetilde{\eta}_j] \exp{ - \int d^dx \ dt \left[ \widetilde{\eta}_j \left( \partial_t \eta^j - B^j_k\eta^k - \xi^j \right) \right]}
    \end{align}
    to get
    \begin{equation}
        \mean{ O[\boldsymbol{\eta}] }_\xi = \int \mathcal{D}[\boldsymbol{\xi}] \prod_j \mathcal{D} [\eta^j] \mathcal{D}[i\widetilde{\eta}_j]\ O[\boldsymbol{\eta}] \exp \left[ \dfrac{1}{4} \int d^dx \ dt \left( \xi^i \Xi^{-1}_ {ij} \xi^j  - 4 \widetilde{\eta}_j \xi^j \right) -  \widetilde{\eta}_j \left( \partial_t \eta^j - B^j_k\eta^k \right) \right].
    \end{equation}
    Completing the square in $\xi$ and doing the Gaussian integrals, we get
    \begin{align}
        \mean{ O[\boldsymbol{\eta}] }_\xi = \int \prod_j \mathcal{D} [\eta^j] \mathcal{D}[i\widetilde{\eta}_ j]\ O[\boldsymbol{\eta}] \times  \exp \left \lbrace - \int d^dx \ dt \left[ \widetilde{\eta}_ j \left( \partial_t \eta^j - B^j_k\eta^k \right) - \widetilde{\eta}_ j \Xi^{jk} \widetilde{\eta}_ k \right] \right\rbrace.
    \end{align}
    Doing the integrals over the response fields $\widetilde{\eta}$, we are left with
    \begin{equation}
        \mean{ O[\boldsymbol{\eta}] }_\xi = \int \prod_j \mathcal{D} [\eta^j] \ O[\boldsymbol{\eta}] \ \exp(-\mathcal{A}[\boldsymbol{\eta}])
    \end{equation}
    where $\mathcal{A}$ is the Onsager-Machlup functional
    \begin{equation}
        \mathcal{A} = -\dfrac{1}{4D} \int d^dx \ dt \ \left( \partial_t \eta^j - B^j_k \eta^k \right)^2
    \end{equation}
    
    Noting that the only time asymmetric part of the action is $\partial_t \eta$, we can write
    \begin{align}
        \mathcal{A} &=  -\dfrac{1}{4D} \int d^dx \ dt \ \left( \partial_t \phi + D\dfrac{\delta F}{\delta \phi} + \alpha \psi \right)^2 + \left( \partial_t \psi + D \dfrac{\delta F}{\delta \psi} - \alpha \phi \right)^2 \\
        \widetilde{\mathcal{A}} &= -\dfrac{1}{4D} \int d^dx \ dt \ \left( \partial_t \phi - D \dfrac{\delta F}{\delta \phi} - \alpha \psi \right)^2 + \left( \partial_t \psi - D \dfrac{\delta F}{\delta \psi} + \alpha \phi \right)^2
    \end{align}
    Taking the difference $\widetilde{\mathcal{A}} - \mathcal{A}$, and noting that $(a+b)^2 - (a-b)^2 = 4ab$, we have
    \begin{equation}
        \widetilde{\mathcal{A}} - \mathcal{A} = -\dfrac{1}{D} \int d^dx dt \ \partial_t \phi \left( -D \dfrac{\delta F}{\delta \phi} - \alpha \psi \right) + \partial_t \psi \left( -D \dfrac{\delta F}{\delta \psi} + \alpha \phi \right)
    \end{equation}
    
    In the Stratonovich convention, $dF/dt = \partial_t \phi \ (\delta F / \delta \phi) + \partial_t \psi \ (\delta F / \delta \psi)$, which will turn into a constant difference in free energies upon taking the time integral. This constant value will tend to zero as the limit $T \to \infty$ is taken. Further, there is a time-symmetric portion of the action that is being omitted due to the Jacobian factor in switching from an integral in $\boldsymbol{\xi}$ to $\boldsymbol{\eta}$ that also arises due to the Stratonovich discretization used throughout this article.
    
    We find the entropy production rate to be
    \begin{equation}
        \label{seq:epr_dgf_realspace}
        \epr = \lim_{T \to \infty} \dfrac{\alpha}{DT} \int d^dx dt \ \left( \psi \dot{\phi} -\dot{\psi} \phi \right)
    \end{equation}
    Plugging in the equations of motion, we find
    \begin{align}
        \mean{ \dot{\psi} \phi } &= \mean{ -D \dfrac{\delta F}{\delta \psi} \phi - \alpha \phi^2 + \xi_\psi \phi } = \mean{ - D \left[ (r - \nabla^2) \psi \right] \phi - \alpha \phi^2 + \xi_\psi \phi } \\
        \mean{ \psi \dot{\phi} } &= \mean{ -D \dfrac{\delta F}{\delta \phi} \psi + \alpha \psi^2 + \xi_\phi \psi } = \mean{ -D \left[ (r - \nabla^2) \phi \right] \psi + \alpha \psi^2 + \xi_\phi \psi }
    \end{align}
    Some care must be taken in evaluating the terms linear in the noise. If the Ito convention had been used, they would be trivially zero, but that is not the general case in the Stratonovich convention. However, as each field is multiplied by the opposite component of the noise, one can show that they indeed identically equal $0$. Putting everything together, we have
    \begin{equation}
        -D r \psi \phi + D (\nabla^2 \psi) \phi + \alpha \phi^2 + D r \phi \psi - D (\nabla^2 \phi) \psi +\alpha \psi^2
    \end{equation}
    The two Laplacian terms will cancel under one integration by parts each, leaving us with
    \begin{equation}
        \epr = \dfrac{\alpha^2}{D} \int d^dx \ \mean{ \phi^2 + \psi^2 }
    \end{equation}
    where we have replaced the time average with an ensemble average, assuming ergodicity. Assuming the system to be in the steady state, we integrate over the equal-time (i.e. $\omega=0$) power spectrum of $\phi$ and $\psi$
    \begin{equation}
        \mean{ \phi(k) \phi(-k) } = \int \dfrac{dk}{2 \pi} \dfrac{1}{k^2 + r} = \dfrac{1}{2\sqrt{r}}.
    \end{equation}
    Using this expression for both $\phi$ and $\psi$ in the equation for $\epr$ above, we have
    \begin{equation}
        \label{seq:gaussfieldEPR_MSR}
        \epr = \dfrac{\alpha^2}{D\sqrt{r}} V,
    \end{equation}
    where $V$ is the total volume of the space, and $\eprfield = \epr / V$.
    \subsection{Generalized Harada-Sasa Relation}
        Here, we show that $\epf^{DGF}$, written in \eq{eq:gaussfieldEPR}, is equivalent to the spectral decomposition based on a generalized Harada-Sasa relation (GHSR) given in \cite{Nardini2017}. The non-equilibrium driving in~\eq{seq:dgf_langevin} comes from a rotational current between the two fields, rather than from a current derived from the gradient of a non-conservative chemical potential. This requires us to derive a slightly altered version of the GHSR than given in \cite{Nardini2017}, largely following the steps outlined therein. In addition, our 2-field problem requires that the response and correlation functions be tensors, and the energy dissipation is related to the trace of their difference \cite{Harada2006EnergySystems}.

        We consider the response of the fields $\boldsymbol{\eta} = \left( \phi, \psi \right)$ to a constant external field, $\left( h_\phi(\mathbf{r}), h_\psi(\mathbf{r}) \right)$, changing the action to
        \begin{equation}
            \mathcal{A}^h =  -\dfrac{1}{4D} \int d^dx \ dt \ \left( \partial_t \phi + D\dfrac{\delta F}{\delta \phi} + D h_\phi + \alpha \psi \right)^2 + \left( \partial_t \psi + D \dfrac{\delta F}{\delta \psi} + D h_\psi- \alpha \phi \right)^2
        \end{equation}
        The response function is upgraded to a response tensor, $\mathcal{R}$, with elements
        \begin{equation}
            R_{ij} \left(\mathbf{r}_{1}, \mathbf{r}_{2}, t\right) = -\bigg\langle\left. \eta_i \left(\mathbf{r}_{1}, t\right) \frac{\delta \mathcal{A}^{h}}{\delta h_j\left(\mathbf{r}_{2}, 0\right)}\right|_{h_j=0} \bigg\rangle
        \end{equation}
        To first order in each external field, the perturbation to the action is
        \begin{equation}
            \delta A^h = -\dfrac{1}{2} \int d^dx dt \ \left[ h_\phi \left( \partial_t \phi + D \dfrac{\delta F}{\delta \phi} + \alpha \psi \right) + h_\psi \left( \partial_t \psi + D \dfrac{\delta F}{\delta \psi} - \alpha \phi \right) \right]
        \end{equation}
        The trace of $\mathcal{R}$, taken at the same spatial location, is
        \begin{align}
            \begin{split}
                \mathrm{Tr}\left(\mathcal{R} (\mathbf{r}, \mathbf{r}, t) \right) = - \dfrac{1}{2} \int d^dx dt \ \Big[ &\phi(\mathbf{r}, t) \left( \partial_t \phi + D \dfrac{\delta F}{\delta \phi} + \alpha \psi \right)\big\vert_{(\mathbf{r}, 0)} \\
                + &\psi(\mathbf{r}, t) \left( \partial_t \psi + D \dfrac{\delta F}{\delta \psi} - \alpha \phi \right)\big\vert_{(\mathbf{r}, 0)} \Big].
            \end{split}
        \end{align}
        The time-asymmetric part of the above equation can be written as
        \begin{align}
            \begin{split}
                \mathrm{Tr}\left[ \mathcal{R}(\mathbf{r}, \mathbf{r}, t) - \mathcal{R}(\mathbf{r}, \mathbf{r}, -t) \right] =& - \mathrm{Tr} \left[ \partial_t C(\mathbf{r}, \mathbf{r}, t)\right] \\
                +& \dfrac{\alpha}{2} \left\langle \left( \phi(\mathbf{r}, t) - \phi(\mathbf{r}, -t)\right) \psi(\mathbf{r}, 0)  - \left( \psi(\mathbf{r}, t) - \psi(\mathbf{r}, -t)\right) \phi(\mathbf{r}, 0)  \right\rangle,
            \end{split}
        \end{align}
        where the correlation matrix has elements $C_{ij}(\mathbf{r}_1, \mathbf{r}_2, t_1-t_2) = \langle \eta_i(\mathbf{r}_1, t_1) \eta_j(\mathbf{r}_2, t_2) \rangle$, equivalent to what was used above to define $\epf$. We have also dropped the terms proportional to $\delta F / \delta \phi$ as they are constants that will be eliminated when we take the time derivative in the next line.
        Looking at \eq{seq:epr_dgf_realspace}, we can rewrite the above as
        \begin{equation}
            \epr = \dfrac{1}{D} \lim_{t \to 0} \int d^dx dt \ \partial_t \mathrm{Tr} \left[ \mathcal{R}(\mathbf{r}, \mathbf{r}, t) - \mathcal{R}(\mathbf{r}, \mathbf{r}, -t) + \partial_t C(\mathbf{r}, \mathbf{r}, t) \right].
        \end{equation}
        Taking the Fourier transform, and denoting the imaginary part of $\mathcal{R}$ as $\widetilde{\mathcal{R}}$, we arrive at our modified GHSR
        \begin{equation}
            \epr =  \int \dfrac{d \omega}{2 \pi} \dfrac{d \mathbf{q}}{(2 \pi)^d} \sigma(\mathbf{q}, \omega); \qquad \sigma(\mathbf{q}, \omega) = \dfrac{\omega}{D} \mathrm{Tr}\left[ \omega C(\mathbf{q}, \omega) - 2\widetilde{\mathcal{R}}(\mathbf{q}, \omega) \right].
        \end{equation}
        \indent It now remains to check if $\sigma = \epf$ for the driven Gaussian fields. As we have already calculated the correlation matrix, we are left to calculate $R_{ij} = \delta \langle \eta_i \rangle / \delta h_j$. Solving the perturbed versions of \eq{seq:gaussfield_eom} in frequency space and taking the mean, we find
        \begin{align}
            \langle \phi \rangle &= \dfrac{D h_\phi + \alpha \langle \psi \rangle}{D(r + q^2) - i \omega} \\
            \langle \psi \rangle &= \dfrac{D h_\psi - \alpha \langle \phi \rangle}{D(r + q^2) - i \omega}.
        \end{align}
        Plugging one solution into the other, we find the auto-responses to be equal to each other, giving
        \begin{equation}
            \mathrm{Tr} \left[ \mathcal{R} \right] = \dfrac{2D\left(D(r + q^2) - i \omega \right)}{\left( D(r + q^2) - i \omega \right)^2 + \alpha^2} \to \mathrm{Tr} \left[ \widetilde{\mathcal{R}} \right] = \dfrac{2D \omega \left( D^2(r + q^2)^2 - \alpha^2 + \omega^2 \right)}{(\omega^2 - \omega_0^2)^2 + (2r\omega)^2},
        \end{equation}
        where $\omega_0$ is defined as in the main text, $\omega_0^2 = D^2(r + q^2)^2 + \alpha^2$. As can be seen in the previous section, the trace of the correlation function is
        \begin{equation}
            \mathrm{Tr}\left[ C \right] = \dfrac{4D \left( D^2(r + q^2)^2 + \alpha^2 + \omega^2 \right)}{(\omega^2 - \omega_0^2)^2 + (2r\omega)^2}.
        \end{equation}
        This finally gets us to our desired result
        \begin{equation}
            \sigma(\mathbf{q}, \omega) = \dfrac{8 \alpha^2 \omega^2}{(\omega^2 - \omega_0^2)^2 + (2r\omega)^2} = \epf^{DGF}.
        \end{equation}
\section{\label{sec:dbp} Driven Brownian particle}

    Here, we consider the particle version of the dynamics given by the Gaussian fields above. We consider an overdamped Brownian particle in a $2$-dimensional harmonic trap with stiffness $r$ subject to a non-conservative, rotational force. This is the particle version of the driven Gaussian fields considered above. The dynamics obey the Langevin equation
    \begin{equation}
        \label{seq:dbp}
        \dot{\mathbf{x}} = A\mathbf{x} + \sqrt{2} \boldsymbol{\xi}; \quad
        A =
        \begin{pmatrix}
            -r & -\alpha &\\
            \alpha & -r 
        \end{pmatrix},
    \end{equation}
    where $\boldsymbol{\xi}$ is Gaussian white noise, $\mean{ \xi_i(t) \xi_j(t') } = \delta_{ij} \delta(t-t')$. These linear dynamics exactly satisfy \eq{eq:gaussPathProb}~\cite{Gardiner2010StochasticSciences}. Using \eq{seq:spectralEPR_particles} with the analytically calculated covariance functions, we find the exact EPF, intregrating to yield the EPR, in agreement with results from stochastic thermodynamics~\cite{Seifert2012}
    \begin{equation}
        \label{eq:dbpEPR}
        \epf^{DBP} = \dfrac{8 \alpha^2 \omega^2}{(\omega^2 - \omega_0^2)^2 + (2 r \omega)^2}, \quad \epr_{thry}^{DBP} = \dfrac{2 \alpha^2}{r}
    \end{equation}
    The covariance functions are calculated in the same way we calculated the covariance functions for the driven Gaussian fields, simply replacing $D(r + q^2) \to r$. For the solution using stochastic thermodynamics, we write the Fokker-Planck equation associated with \eq{seq:dbp} and solving for the steady state probability density, $p_{ss}$ as well as the steady state current, $j_{ss}$, given by
    \begin{equation}
        p_{ss}(\mathbf{x}) = \dfrac{r}{2 \pi} e^{-r x^2 /2}; \qquad j_{ss}(\mathbf{x}) = \alpha \mathbf{x} p^{ss}.
    \end{equation}
    We can then calculate $\epr$ as~\cite{Seifert2012}
    \begin{equation}
        \dot{S} = \int d \mathbf{x} \ \dfrac{j_{ss}^2}{p_{ss}} = \dfrac{2 \alpha^2}{r}
    \end{equation}
    
    Similarly to $\epffield^{DGF}$, $\epf^{DBP}$ is peaked at $\omega_0 = (r^2 + \alpha^2)^{1/2}$ and decays as $\omega^2$ for large $\omega$. While multiple combinations of $\alpha$ and $r$ can give the same value for $\epr$, $\epf$ distinguishes between equally dissipative trajectories in the shape and location of its peaks, giving information about the form of the underlying dynamics while retaining the same total EPR.

    To test whether we would be able to infer these functions from data, we generate trajectories of length $T$ [See methods for details] and estimate $\epf$ from the trajectories alone. We see an excellent agreement between the measured $\smooth{\epf}$ and \eq{eq:dbpEPR}, while also maintaining agreement with $\epr^{DBP}$ upon numerical integration of $\smooth{\epf}$ (\sfig{sfig:eprPlot_dbp}). This remains true for 3, and 4 dimensional simulations (\sfig{sfig:dbp_234dim}), highlighting our ability to estimate $\epf$ for high dimensional data.

\section{Macroscopic Brusselator Dynamics}
    The reversible Brusselator model we consider in this paper is defined by
    
    \begin{equation}
    \label{seq:brusselator}
        A \underset{k^-_1}{\overset{k^+_1}{\rightleftharpoons}} X; \quad
        B + X \underset{k^-_2}{\overset{k^+_2}{\rightleftharpoons}} Y + C; \quad
        2X + Y \underset{k^-_3}{\overset{k^+_3}{\rightleftharpoons}} 3X,
    \end{equation}
    where $A, B, C$ are fixed external chemicals and the system is assumed to occur in a well-mixed vessel of volume $V$. Using mass action kinetics and writing lower-case letters as concentrations (e.g. $x \equiv X / V$), the macroscopic dynamics of the Brusselator are given by the coupled ODEs
    \begin{align}
    \label{seq:bruss_macroEOMs}
        \begin{split}
            \dot{x} &= k_1^+ a - k_1^- x - k_2^+ b x + k_2^- c y + k_3^+ x^2 y - k_3^- x^3 \\
            \dot{y} &= k_2^+ b x - k_2^- c y - k_3^+ x^2 y + k_3^- x^3
        \end{split}
    \end{align}

    Detailed balance holds when each reaction rate in \eq{seq:brusselator} is balanced, leading to the following equilibrium concentrations
    \begin{align}
        X_{eq} &= A \dfrac{k_1^+}{k_1^-} \\
        Y_{eq} &= X_{eq} \dfrac{B k_2^+}{C k_2^-} = X_{eq} \dfrac{k_3^-}{k_3^+},
    \end{align}
    where the first and second equation for $y_{eq}$ come from the $k_2$ and $k_3$ reactions, respectively. Using the two equations for $y_{eq}$ gives us the condition for detailed balance given in the main text, $B k_2^+ k_3^+ = C k_2^- k_3^-$.
    
    The steady state values of $(x, y)$ are given by setting the deterministic equations to $0$, giving
    \begin{align}
        x_{ss} &= a \dfrac{k_1^+}{k_1^-} \\
        y_{ss} &= \dfrac{k_2^+ b x_{ss} + k_3^- x_{ss}^3}{k_2^- c + k_3^+ x_{ss}^2}
    \end{align}
    The relaxation matrix, $R$, that defines the stability of the steady state is given by expanding the deterministic equations to first order around their steady state values
    \begin{equation}
        R =
        \begin{pmatrix}
            \partial_x \dot{x} & \partial_y \dot{x} \\
            \partial_x \dot{y} & \partial_y \dot{y} \\
        \end{pmatrix}\bigg\vert_{ss}
        =
        \begin{pmatrix}
            -(k_1^- + b k_2^+) + 2 k_3^+ x_{ss} y_{ss} - 3 k_3^- x_{ss}^2 & k_2^- c + k_3^+ x_{ss}^2 \\
            b k_2^+ - 2 k_3^+ x_{ss} y_{ss} + 3 k_3^- x_{ss}^2 & -(k_2^- c + k_3^+ x_{ss}^2)
        \end{pmatrix}.
    \end{equation}
    The eigenvalues of $R$ are given by solving its characteristic equation, giving
    \begin{equation}
        \lambda_\pm = \dfrac{\mathrm{Tr}(R)}{2} \pm \left[ \left(\dfrac{\mathrm{Tr}(R)}{2} \right)^2 - \det(R) \right]^{1/2}
    \end{equation}

\section{Alternative $\epr$ estimators}
\begin{figure*}[ht]
    \centering
    \includegraphics[width=0.75\textwidth]{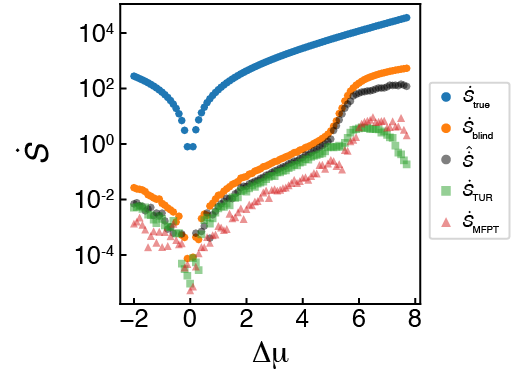}
    \caption{(left) Comparison of $\smooth{\epr}$ and $\epr_{v^2}$, showing qualitatively similar results, indicating that $\smooth{\epr}$ contains the same information as a Fokker-Planck description of the system. (right) Same plot on a log-log scale. The red dashed lines have a slope of 2, the reflecting the fact that the entropy production rate near equilibrium is quadratic in the driving force $\Delta \mu$. While $\smooth{\epr}$ is quadratic, $\epr_{v^2}$ is not, showing us that $\smooth{\epr}$ is a more accurate measure of $\epr_{true}$ near equilibrium. \textbf{Alternative methods for measuring $\epr$.} Comparison of $\smooth{\epr}$ (black dots, same data as in \fig{fig:epr_brusselator}C) with two alternative methods for estimating entropy production rates. $\dot{S}_\mathrm{TUR}$ (green squares) is based on the thermodynamic uncertainty relation (TUR), and $\dot{S}_\mathrm{MFPT}$ is based on measuring the mean first passage time of an observable. All estimates approximate $\dot{S}_\mathrm{blind}$ because they are based only on observables in the $(X, Y)$ plane. Our estimator, $\smooth{\epr}$, outperforms the other two estimators, especially beyond the Hopf bifurcation. See Supplementary Materials for details regarding the implementation of the TUR and MFPT methods.}
    \label{sfig:eprPlot_bruss_altMethods}
\end{figure*}

    Supplementary Figure \ref{sfig:eprPlot_bruss_altMethods} shows our original results in \fig{fig:epr_brusselator} in addition to results from two other estimators of $\epr$. Here, we describe these alternative methods and our implementation of them.
    
    One of the estimators is based on the thermodynamic uncertainty relation (TUR) \cite{Barato2015ThermodynamicProcesses, Horowitz2017}, which we implemented following \cite{Li2019QuantifyingCurrents}. To summarize the method, empirical phase space fluxes, $\mathbf{j}(\mathbf{x}, t)$ are integrated over space with a weighting vector field $\mathbf{d}(\mathbf{x})$ to give a macroscopic current, $j_\mathbf{d}$, that accumulates over an observation time:
    \begin{equation}
        j_\mathbf{d}(\tau_\mathrm{obs}) = \int\limits_0^{\tau_\mathrm{obs}} dt \int d\mathbf{x} \ \mathbf{j}(\mathbf{x}, t) \cdot \mathbf{d}(\mathbf{x}).
    \end{equation}
    The mean and variance of $j_\mathbf{d}$ are then used to estimate $\epr$ as
    \begin{equation}
        \epr \geq \epr_\mathrm{TUR} = \dfrac{2 \langle j_\mathbf{d} \rangle}{\tau_\mathrm{obs} \mathrm{Var}(j_\mathbf{d})}.
    \end{equation}
    The choice of weighting field $\mathbf{d}$ was shown in \cite{Li2019QuantifyingCurrents} to significantly affect the tightness of the bound given above. While they devised a Monte-Carlo procedure to minimize the mean-square error of $\epr_\mathrm{TUR}$, we opted to use a choice of $\mathbf{d}$ informed by the knowledge of the underlying dynamics. Namely, we chose $\mathbf{d} = \left(\dot{x}, \dot{y} \right)$, defined in \eq{seq:bruss_macroEOMs}.
    
    The other estimator is based on the mean first passage time (MFPT) of an observable, $\mathcal{O}$, based on the method presented in \cite{Roldan2015}. The method involves measuring the average time, $\langle \tau_\mathcal{O} \rangle$, it takes for $\mathcal{O}$ to reach a threshold value $L(\alpha) = \ln((1 - \alpha) / \alpha)$, where $\alpha$ is the fraction of false-positives (and false-negatives) deemed tolerable by the user. $\epr$ is then bounded by:
    \begin{equation}
        \epr \geq \dfrac{L(\alpha) (1 - 2\alpha)}{\langle \tau_\mathcal{O} \rangle}.
    \end{equation}
    If $\mathcal{O}$ is chosen to be the log-likelihood ratio between the probabilities of observing a given time series conditioned on the hypothesis that the time series is played forwards or backwards, $\langle \tau_\mathcal{O} \rangle$ is minimized and the above bound is saturated. In order to restrict ourselves to dynamics observable in the $XY$ plane, we choose $\mathcal{O}$ based on the winding number of the trajectory around the trajectory's center of mass, $w(t) = \theta(t) / 2\pi$, where $\theta(t) = \arctan((Y - \langle Y \rangle) / (x - \langle X \rangle))$, and $\theta(t)$ is measured cumulatively (e.g. two full counter-clockwise rotations give $\theta = 4\pi$). We then make the assumption that $w$ is a drift-diffusion process that obeys the Langevin equation $\dot{w} = v + \sqrt{2D}\xi$, where $\xi$ is unit variance Gaussian noise. The optimal observable for measuring the MFPT in this case is given by $\mathcal{O}(t) = \frac{v}{D}\left( w(t) - w(0) \right)$, where $v = \langle w(T) \rangle / T$ and $D = \mathrm{Var}(w(T))/2T$, $T$ is the total observation time, and the mean and variance are calculated over many realizations of the same dynamics.

\section{Simulation Parameters}
    \subsection{Gaussian Fields}
        Simulations of the Gaussian fields use an Euler Maruyama algorithm to integrate the equations of motion. Time and space are scaled by $\tau = (Dr)^{-1}$ and $\lambda = r^{-1/2}$. The simulation is performed on a periodic, 1 dimensional lattice.
        
        \begin{center}
        \begin{tabular}{ |c|c|c|l| }
        \hline
            \multicolumn{4}{|c|}{Parameters} \\
            \hline
             Name & Value & units & Description \\
             \hline
             $N_{sim}$ & 10 & 1 & number of simulations per parameter set \\ 
             $dt$ & 0.0001 & $\tau$ & simulation time step \\
             $t_{final}$ & 50 & $\tau$ & total simulation time \\
             $dx$ & 0.1 & $\lambda$ & spacing between lattice sites \\
             $N_{sites}$ & 128 & 1 & number of lattice sites \\
             $\alpha$ & $\left[ 0, 1, \ldots, 25 \right]$ & $\tau^{-1}$ & driving frequency \\
             $\sigma_\omega$ & 1.57 & $\tau^{-1}$ & width of Gaussian used to smooth in the temporal dimension \\
             $\sigma_k$ & 1.47 & $\lambda^{-1}$ & width of Gaussian used to smooth in the spatial dimension \\
        \hline
        \end{tabular}
        \end{center}
    
    \subsection{Brusselator}
        Simulations of the Brusselator are done using a Gillespie algorithm~\cite{Gillespie1977}. Time is non-dimensionalized by $\tau = 1 / k_1^+$. 
        
        \begin{center}
        \begin{tabular}{ |c|c|c|l| }
        \hline
            \multicolumn{4}{|c|}{Parameters} \\
            \hline
             Name & Value & units & Description \\
             \hline
             $N_{sim}$ & 50 & 1 & number of simulations per parameter set \\
             $t_{final}$ & 5000 & $\tau$ & total time of simulation \\
             $k_1^+$ & 1 & $\tau^{-1}$ & forward reaction rate for reaction 1 \\
             $k_1^-$ & 0.5 & $\tau^{-1}$ & reverse reaction rate for reaction 1 \\
             $k_2^+$ & 2 & $\tau^{-1}$ & forward reaction rate for reaction 2 \\
             $k_2^-$ & 0.5 & $\tau^{-1}$ & reverse reaction rate for reaction 2 \\
             $k_3^+$ & 2 & $\tau^{-1}$ & forward reaction rate for reaction 3 \\
             $k_3^-$ & 0.5 & $\tau^{-1}$ & reverse reaction rate for reaction 3 \\
             $A$ & 100 & 1 & number of chemical species in reaction volume \\
             $V$ & 100 & 1 & reaction volume, used for calculated propensities in Gillespie algorithm \\
        \hline
        \end{tabular}
        \end{center}
        The strength of external driving is given by $\Delta \mu = \ln \left( (b k_2^+ k_3^+)/(c k_2^- k_3^-) \right)$, where $b = B/V$ and $c = C/V$. Values of $B$ and $C$ are changed to give driving strengths $\Delta \mu \in [-2, 8]$ with step size 0.1, while keeping the product $\sqrt{(B k_2^+ k_3^+)(C k_2^- k_3^-)} = 1,000$, where 1,000 $\sqrt{(b k_2^+ k_3^+)(c k_2^- k_3^-)} = 1$, where 1 is an arbitrarily chosen constant. The EPR plot in \fig{fig:epr_brusselator}(c) uses varying smoothing widths.
        \begin{itemize}
            \item $\Delta \mu < 5 \to \sigma = 1.26$
            \item $\Delta \mu \in [5, 5.8] \to \sigma = 0.063$
            \item $\Delta \mu > 5.8 \to \sigma = 0.031$
        \end{itemize}
        The EPF plot shown in \fig{fig:epr_brusselator}b uses a smoothing width of $\sigma = 0.126$. Different system volumes are used in \fig{fig:bruss_eprScaling}.

    \subsection{Reaction-diffusion Brusselator}
        To add reactions to the Brusselator, we employ a compartment-based Gillespie algorithm. We non-dimensionalize time by $\tau = 1/k_1^+$ and use the distance between each compartment as our length scale $\lambda = h = 1$.
        
        \begin{center}
        \begin{tabular}{ |c|c|c|l| }
        \hline
            \multicolumn{4}{|c|}{Parameters} \\
            \hline
             Name & Value & units & Description \\
             \hline
             $N_{sim}$ & 10 & 1 & number of simulations per parameter set \\
             $N_c$ & 64 & 1 & number of lattice sites \\
             $D_X$ & 1 & $\lambda^2 \tau^{-1}$ & diffusion constant of chemical species $X$ \\
             $D_Y$ & 0.1 & $\lambda^2 \tau^{-1}$ & diffusion constant of chemical species $Y$ \\
             $t_{final}$ & 100 & $\tau$ & total time of simulation \\
             $k_1^+$ & 1 & $\tau^{-1}$ & forward reaction rate for reaction 1 \\
             $k_1^-$ & 0.5 & $\tau^{-1}$ & reverse reaction rate for reaction 1 \\
             $k_2^+$ & 2 & $\tau^{-1}$ & forward reaction rate for reaction 2 \\
             $k_2^-$ & 0.5 & $\tau^{-1}$ & reverse reaction rate for reaction 2 \\
             $k_3^+$ & 2 & $\tau^{-1}$ & forward reaction rate for reaction 3 \\
             $k_3^-$ & 0.5 & $\tau^{-1}$ & reverse reaction rate for reaction 3 \\
             $A$ & 100 & 1 & number of chemical species in reaction volume \\
             $C$ & 400 & 1 & number of chemical species in reaction volume \\
             $V$ & 100 & 1 & reaction volume of each compartment, used to calculated propensities \\
        \hline
        \end{tabular}
        \end{center}
        
        The strength of the driving $\Delta \mu$, and therefore the values of $B$ and $C$, is calculated in the same way as done for the Brusselator simulations without reactions. All subpanels involving the reaction-diffusion Brusselator use the same parameters with $\Delta \mu$ specified in the figure caption. \fig{fig:epr_brussfield} uses a varying smoothing width
        \begin{itemize}
            \item $\Delta \mu \in [-1, -0.5] \to (\sigma_\omega, \sigma_q) = (7, 1)$
            \item $\Delta \mu \in [-0.5, 0.5] \to (\sigma_\omega, \sigma_q) = (14, 2)$
            \item $\Delta \mu \in [0.5, 1] \to (\sigma_\omega, \sigma_q) = (7, 1)$
            \item $\Delta \mu \in [1, 5.65] \to (\sigma_\omega, \sigma_q) = (5, 0.5)$
            \item $\Delta \mu = 5.7 \to (\sigma_\omega, \sigma_q) = (3.5, 0.4)$
            \item $\Delta \mu = 5.8 \to (\sigma_\omega, \sigma_q) = (1.75, 0.3)$
            \item $\Delta \mu = 5.9 \to (\sigma_\omega, \sigma_q) = (0.7, 0.1)$
            \item $\Delta \mu = 6.0 \to (\sigma_\omega, \sigma_q) = (0.35, 0.05)$
            \item $\Delta \mu = 6.1 \to (\sigma_\omega, \sigma_q) = (0.14, 0.02)$
            \item $\Delta \mu \geq 6.2 \to (\sigma_\omega, \sigma_q) = (0.035, 0.005)$
        \end{itemize}
        
        and \fig{fig:epfPlot_brussfield} use a smoothing width of $(\sigma_\omega, \sigma_q) = (0.07, 0.1)$.
        
    \subsection{Driven Brownian Particle}
        Simulations of the driven Brownian particle shown in the Supplementary Material use an Euler-Maruyama algorithm to integrate the equations of motion. Time is scaled by $\tau = \gamma/k$, which also scales the driving force. The variables have units of length and are scaled by $\lambda = \sqrt{D \gamma /k}$.

\begin{figure*}[ht]
    \centering
    \includegraphics[width=\textwidth]{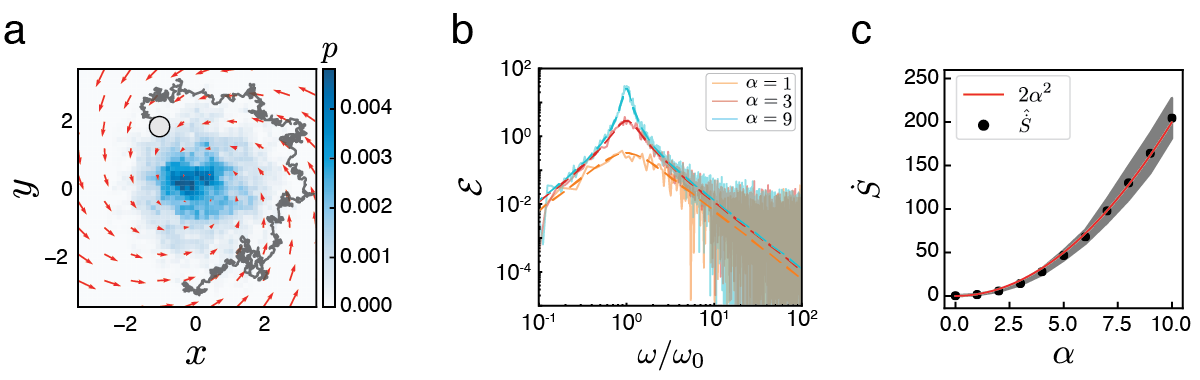} \caption{\textbf{$\epr$ of a driven Brownian particle.} (a) Sample trajectory for simulation of driven Brownian particle in 2 dimensions with $\alpha = 2$ in nondimensionalized units shown in gray with end of trajectory shown with white circle. The heatmap gives the empirical steady-state probability distribution function of particle positions and the red arrows indicate the underlying force field $F(\mathbf{x})$. (b) $\epf$ for $\alpha = [1, 3, 9]$ measured from simulations and calculated from \eq{eq:dbpEPR}, shown in solid and dashed lines respectively. $\epf$ is symmetric in $\omega$, so only the positive axis is shown. (c) EPR for $\alpha = \{0,1,\ldots, 10\}$, smoothed by a Gaussian with $\sigma_\omega=2.1$. Mean $\pm$ standard deviation of $\smooth{\epr}$ over $N=64$ simulations shown with black dots and shaded area. Red line shows non-dimensionalized theoretical value of $\epr$. See Supplementary Materials for all simulation parameters.}
    \label{sfig:eprPlot_dbp}
\end{figure*}

\begin{figure*}[ht]
    \centering
    \includegraphics[width=0.9\textwidth]{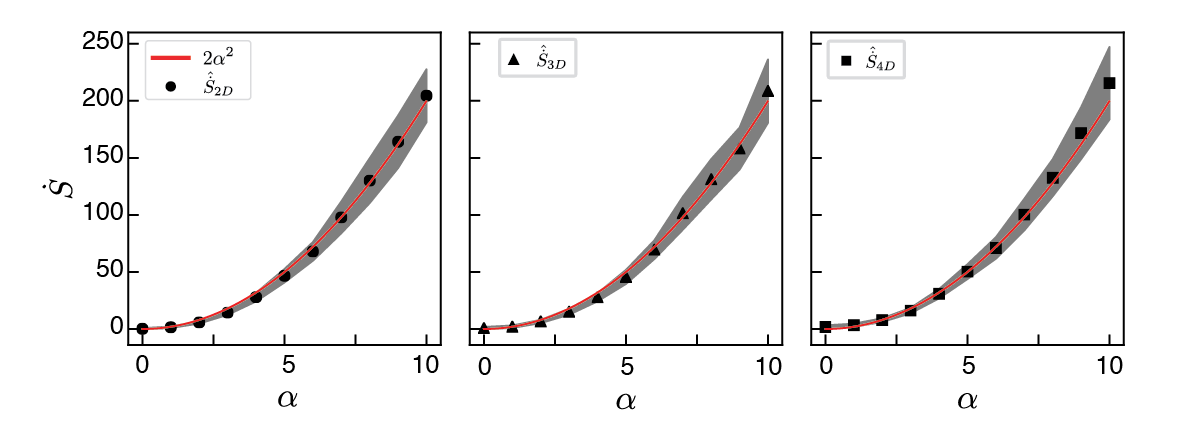}
    \caption{\textbf{$\epr$ of driven Brownian particle in higher dimensions.} $\smooth{\epr}$ for driven Brownian particle simulation in $d = [2, 3, 4]$. Additional dimensions only contain a harmonic potential term, maintaining the non-equilibrium driving only in the $2^\mathrm{nd}$ dimension. All other parameters the same as in \sfig{sfig:eprPlot_dbp}.}
    \label{sfig:dbp_234dim}
\end{figure*}

        \begin{center}
        \begin{tabular}{ |c|c|c|l| }
        \hline
            \multicolumn{4}{|c|}{Parameters} \\
            \hline
             Name & Value & units & Description \\
             \hline
             $N_{sim}$ & $64$ & $1$ & number of simulations per parameter set \\ 
             $N_{dim}$ & $2$  & $1$ & number of dimensions of the simulation \\ 
             $dt$ & $0.001$ & $\tau$ & simulation time step \\ 
             $t_{final}$ & 100 & $\tau$ & total simulation time \\
             $\alpha$ & $\left[0, 1, \ldots, 10\right]$ & $\tau^{-1}$ & strength of driving \\
        \hline
        \end{tabular}
        \end{center}
        \sfig{sfig:eprPlot_dbp} and \fig{sfig:dbp_234dim} use a smoothing width of $\sigma= 1.88$.

\vfill
\pagebreak
\section{Supplementary Figures}

\begin{figure*}[ht]
    \centering
    \includegraphics[width=\textwidth]{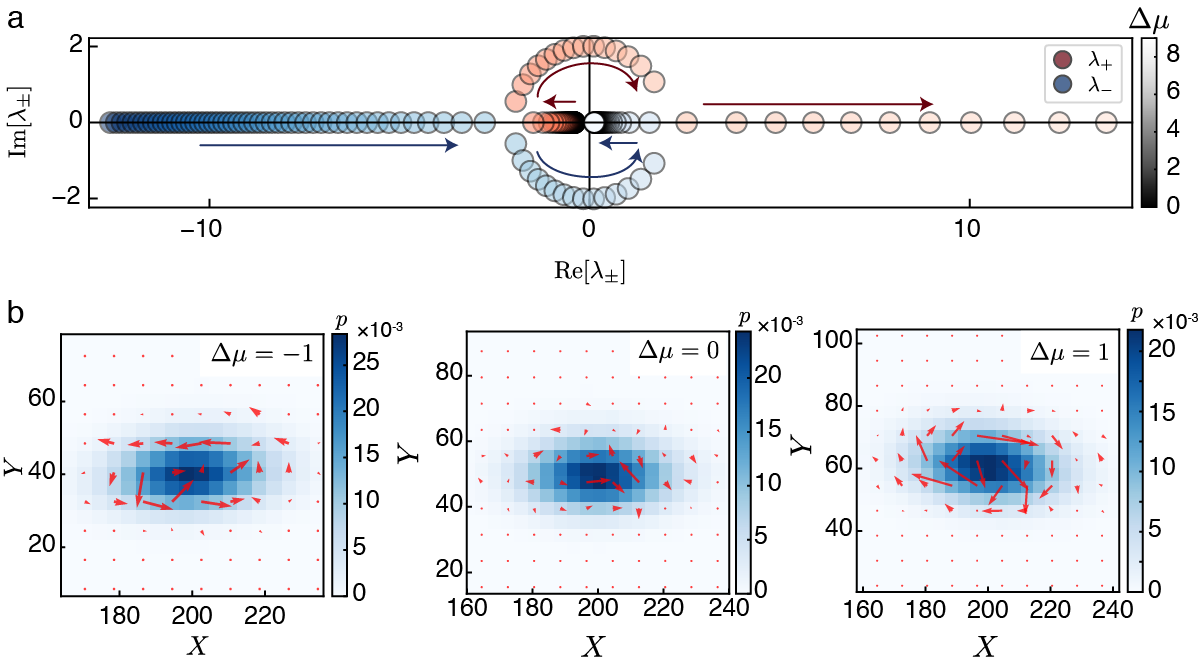}
    \caption{\textbf{Brusselator dynamics exhibit circulation without macroscopic oscillatory solution.} (a) Eigenvalues of the Brusselator's relaxation matrix, $R$ as a function of $\Delta \mu$. $\lambda_\pm$ shown in red and blue, respectively, with each color going from dark to light with increasing $\Delta \mu$. The red and blue arrows serve as guides for the reader to follow the trajectory of $\lambda_\pm$. With our parameters, the stable focus appears at $\Delta \mu = 5.26$ and the Hopf bifurcation occurs at $\Delta \mu_{\text{HB}} = 6.16$. (b) Probability distributions (blue) and probability fluxes (red arrows) for Brusselator simulations with $\Delta \mu = [-1, 0, 1]$, showing the reversal in flux circulation direction at $\Delta \mu = 0$.}
    \label{sfig:brusselator_evalTraj}
\end{figure*}

\begin{figure*}[ht]
    \centering
    \includegraphics[width=0.5\textwidth]{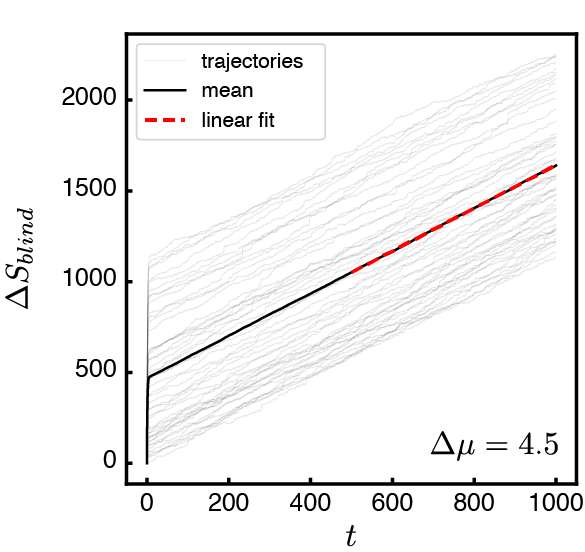}
    \caption{\textbf{Calculating $\epr_\mathrm{true}$ and $\epr_\mathrm{blind}$} Fit to (blinded) entropy produced for Brusselator. Light gray lines show the amount of entropy produce as a function of simulation time for $N=50$ simulations at $\Delta \mu = 4.5$. Each simulation starts at a random initial condition and rapidly approaches the steady state value for $(X,Y)$. This transient trajectory results in the large variation in initial entropy production which depends on how far the system begins from $(X_{ss}, Y_{ss})$. Once the system reaches its steady state, the rate of entropy production approaches a steady value. The average of $\Delta S$ is taken across all trajectories, and a linear fit to the second half of the resulting mean gives us our value of $\epr_{blind}$ given in \fig{fig:epr_brusselator}b. The same method is used to calculate $\epr_{true}$ as well as $\eprfield_{blind}$ and $\eprfield_{true}$ for the reaction-diffusion Brusselator model in \fig{fig:epr_brussfield}.}
    \label{sfig:brussFit}
\end{figure*}

\begin{figure*}[ht]
    \centering
    \includegraphics[width=0.5\textwidth]{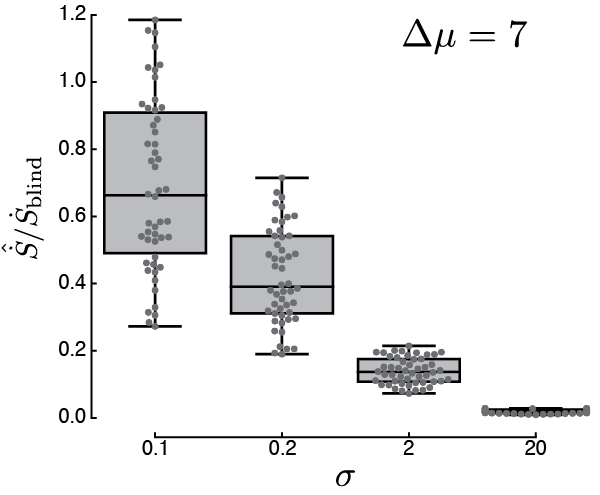}
    \caption{\textbf{Effect of increasing smoothing width.} Underestimation of $\epr_{\mathrm{blind}}$ as a function of smoothing width, $\sigma$, for $\Delta \mu = 7$. We see a systematic decrease in the estimated EPR, $\smooth{\epr}$, as the smoothing width gets wider. Individual points show results for each of the $N=50$ simulations. Center line shows the median, edges of the box show interquartile range, and whiskers show range of data.}
    \label{sfig:eprPlot_bruss_varySigma}
\end{figure*}

\begin{figure*}[ht]
    \centering
    \includegraphics[width=0.75\textwidth]{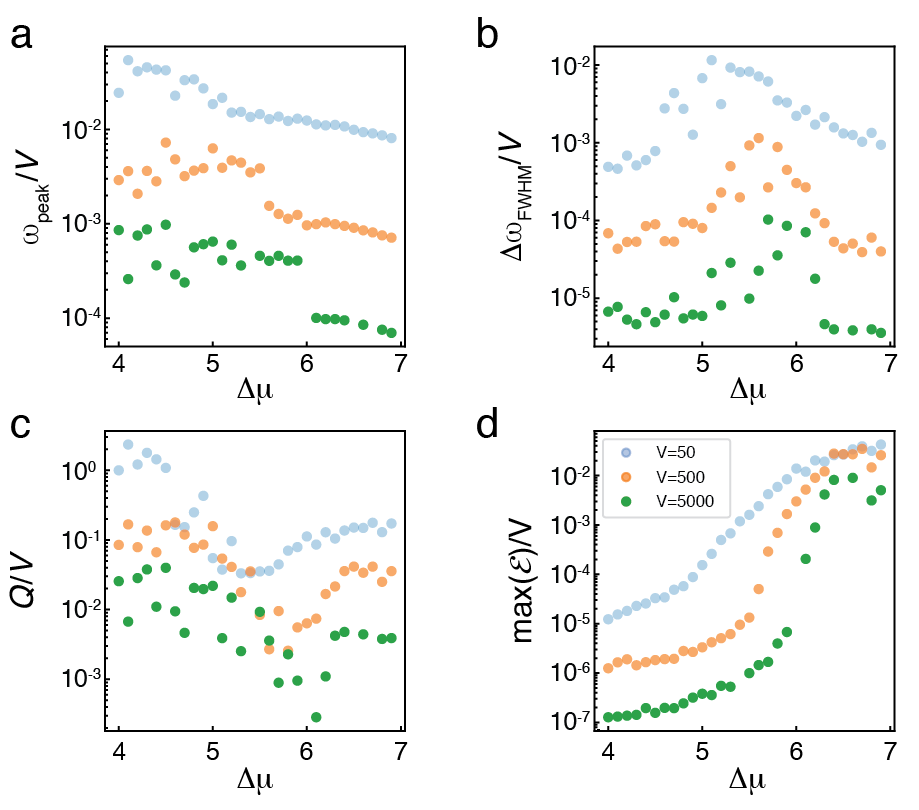}
    \caption{\textbf{Finite size scaling of $\epf$ of Brusselator.} (a) Normalized frequency of maximum of $\epf$, $\omega_\mathrm{peak}/V$ is independent of $V$, but the jump from high to low frequency occurs more sharply and occurs closer to $\Delta \mu_\mathrm{HB}$ as $V$ increases. (b) Normalized full-width half-maximum (FWHM) of peak in $\epf$, $\Delta \omega_\mathrm{FWHM}$, is independent of $V$ and is maximized around the transition point, reflecting the increased fluctuations near the phase transition. The location of the peak moves closer to $\Delta \mu_\mathrm{HB}$ as $V$ increases. (c) The normalized quality factor of $\epf$, $Q/V = \omega_\mathrm{peak} / \Delta \omega_\mathrm{FWHM}V$, is independent of system size, and has a minimum at the transition point. (d) The normalized maximum value of $\epf$ is independent of $V$ below the transition, and gains a linear dependence on $V$ above it, similar to $\epr_\mathrm{blind}$.}
    \label{sfig:bruss_epfscaling}
\end{figure*}

\begin{figure*}[ht]
    \centering
    \includegraphics[width=0.5\textwidth]{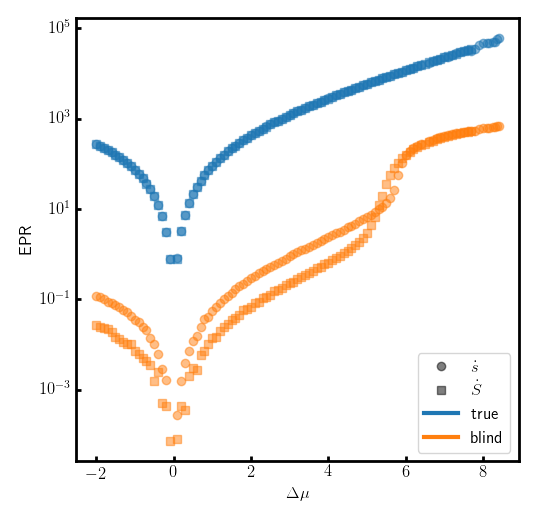}
    \caption{\textbf{Comparing $\epr$ and $\eprfield$ for the well-mixed and reaction-diffusion Brusselator.} True (blue) and blind (orange) entropy production rates for the well-mixed Brusselator ($\epr$, squares) and reaction-diffusion Brusselator ($\eprfield$, circles). These are the same data shown in \fig{fig:epr_brusselator}c and \fig{fig:epr_brussfield}c, plotted together. $\epr_\mathrm{true} \approx \eprfield_\mathrm{true}$ for all driving forces $\Delta \mu$. By contrast, $\eprfield_\mathrm{blind} > \epr_\mathrm{blind}$ below $\Delta \mu_\mathrm{HB}$ due to additional irreverisibility from diffusion between neighboring lattice sites with different concentrations of $(X, Y)$ due to the incoherent dynamics, and $\epr_\mathrm{blind} \approx \eprfield_\mathrm{blind}$ above $\Delta \mu_\mathrm{HB}$ due to the synchronized oscillations, making diffusion between lattice sites an equilibrium process.}
    \label{sfig:bruss_vs_brussfield_EPR}
\end{figure*}

\begin{figure*}[ht]
    \centering
    \includegraphics{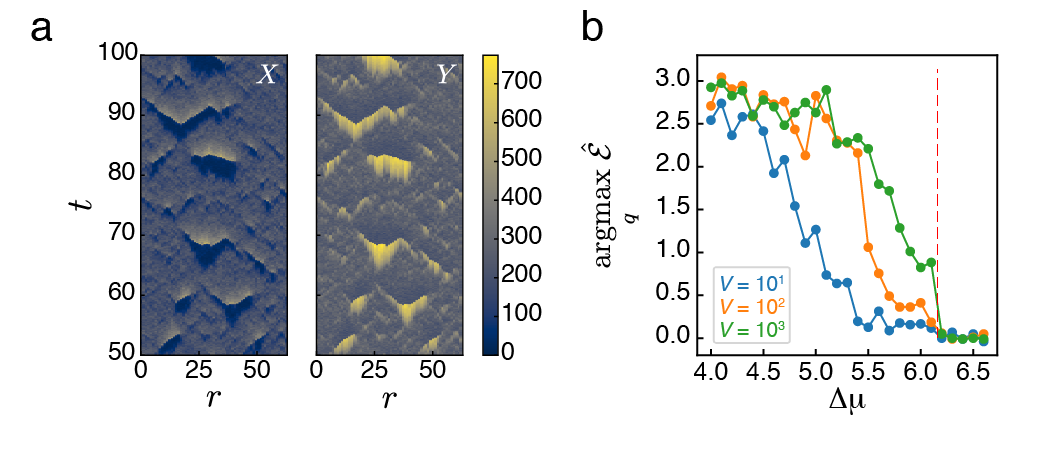}
    \caption{\textbf{Transiently synchronized dynamics in the reaction-diffusion Brusselator and finite-size scaling in $\epffield$. }(a) Typical trajectory of a reaction-diffusion Brusselator system just below the Hopf bifurcation, at $\Delta \mu = 5.8$ and $V=100$. See some flashes on collective behavior, but it does not span the entire system, showing why $\epf$ has peaks somewhere between $q=0$ and the maximum in \fig{fig:epfPlot_brussfield}(c). (b) Wavenumber $q$ that maximizes $\smooth{\epf}$ for the reaction-diffusion Brusselator for compartment volumes $V = \lbrace 10^1, 10^2, 10^3 \rbrace$ shows a sharper transition that gets closer to $\Delta \mu_\mathrm{HB}$ (red line) as the volume increases.}
    \label{sfig:brussfield_preHopfTraj}
\end{figure*}

\end{document}